\pdfoutput=1
\documentclass[twocolumn]{aastex63}
\usepackage[utf8x]{inputenc}
\usepackage{graphicx}
\usepackage{epstopdf}
\usepackage{hyperref}
\usepackage{xcolor}
\usepackage{wrapfig}
\usepackage[utf8x]{inputenc}
\usepackage{rotating}
\usepackage{natbib}
\usepackage{mathtools}
\usepackage[varg]{txfonts}

\begin{document}

\author[0000-0002-8053-8040]{Marianna Annunziatella}
\affil{Centro de Astrobiolog\'ia, (CAB, CSIC-INTA), Carretera de Ajalvir km 4, E-28850 Torrejón de Ardoz, Madrid, Spain}

\author[0000-0002-1917-1200]{Anna Sajina}
\affil{Tufts University, Medford, MA, USA}

\author[0000-0001-7768-5309]{Mauro Stefanon}
\affil{Leiden Observatory, Leiden University, PO Box 9513, NL-2300 RA, Leiden, The Netherlands}
\affiliation{Departament d'Astronomia i Astrof\'isica, Universitat de Val\`encia, C. Dr. Moliner 50, E-46100 Burjassot, Val\`encia,  Spain}

\author[0000-0001-9002-3502]{Danilo Marchesini}
\affil{Tufts University, Medford, MA, USA}

\author[0000-0002-3032-1783]{Mark Lacy}
\affil{NRAO, Charlottesville, VA, USA}

\author[0000-0002-2057-5376]{Ivo Labb\'e}
\affil{Centre for Astrophysics and Supercomputing, Swinburne University of Technology, Australia}

\author{Lilianna Houston}
\affil{Tufts University, Medford, MA, USA}

\author[0000-0001-5063-8254]{Rachel Bezanson}
\affiliation{Department of Physics and Astronomy and PITT PACC, University of Pittsburgh, Pittsburgh, PA 15260, USA}

\author[0000-0003-1344-9475]{Eiichi Egami}
\affil{University of Arizona, USA}

\author[0000-0003-3310-0131]{Xiaohui Fan}
\affil{University of Arizona, USA}

\author[0000-0003-1748-2010]{Duncan Farrah}
\affil{Institute for Astronomy, University of Hawaii, 2680 Woodlawn Dr., Honolulu, HI, 96822, USA}
\affil{Department of Physics and Astronomy, University of Hawai‘i at Mānoa, 2505 Correa Rd., Honolulu, HI, 96822, USA}

\author[0000-0002-5612-3427]{Jenny Greene}
\affil{Princeton, NJ, USA}

\author[0000-0003-4700-663X]{Andy Goulding}
\affil{Princeton, NJ, USA}

\author[0000-0002-1332-5475]{Yen-Ting Lin}
\affil{Institute of Astronomy and Astrophysics, Academia Sinica, Taipei 10617, Taiwan}

\author[0000-0003-0049-5210]{Xin Liu}
\affiliation{Department of Astronomy, University of Illinois at Urbana-Champaign, Urbana, IL 61801, USA}
\affiliation{National Center for Supercomputing Applications, University of Illinois at Urbana-Champaign, Urbana, IL 61801, USA}

\author[0000-0002-3305-9901]{Thibaud Moutard}
\affil{Institute for Computational Astrophysics and Department
of Astronomy and Physics, Saint Mary’s University, 923
Robie Street, Halifax, Nova Scotia, B3H 3C3, Canada}
\affil{Aix Marseille Universite, CNRS, CNES, LAM - Laboratoire d’Astrophysique de Marseille, 38 rue F. Joliot-Curie, F-13388 Marseille, France}

\author[0000-0001-9011-7605]{Yoshiaki Ono}
\affil{University of Tokyo, Japan}

\author[0000-0002-1049-6658]{Masami Ouchi}
\affil{University of Tokyo, Japan}

\author[0000-0002-7712-7857]{Marcin Sawicki}
\affil{Institute for Computational Astrophysics and Department
of Astronomy and Physics, Saint Mary’s University, 923
Robie Street, Halifax, Nova Scotia, B3H 3C3, Canada}

\author[0000-0001-7291-0087]{Jason Surace}
\affil{Infrared Processing and Analysis Center, Pasadena, CA, USA}

\author[0000-0001-7160-3632]{Katherine Whitaker}
\affil{UMass Amherst, Amherst, MA, USA}

\title{The Spitzer Coverage of HSC-Deep with IRAC for Z studies (SHIRAZ) I: IRAC mosaics}

\begin{abstract}
We present new {\it Spitzer} Infrared Array Camera (IRAC) 3.6 and 4.5$\mu$m mosaics of three fields, E-COSMOS, DEEP2-F3, and ELAIS-N1. Our mosaics include both new IRAC observations as well as re-processed archival data in these fields. These fields are part of the HSC-Deep $grizy$ survey and have a wealth of additional ancillary data. The addition of these new IRAC mosaics is critical in allowing for improved photometric redshifts and stellar population parameters at cosmic noon and earlier epochs. The total area mapped by this work is $\sim$ 17 $\mathrm{deg^2}$ with a mean integration time of $\approx$1200s, providing a median 5$\sigma$ depth of 23.7(23.3) at 3.6(4.5)$\mu$m in AB. We perform SExtractor photometry both on the combined mosaics as well as the single-epoch mosaics taken $\approx$6 months apart. The resultant IRAC number counts show good agreement with previous studies.  In combination with the wealth of existing and upcoming spectro-photometric data in these fields, our IRAC mosaics will enable a wide range of galactic evolution and AGN studies. With that goal in mind, we make the combined IRAC mosaics and coverage maps of these three fields publicly available.

counts show good agreement with previous studies.  
\end{abstract}


\graphicspath{{./}}

\section{Introduction}

Over the last three decades, multi-wavelength extragalactic surveys have transformed our understanding of how galaxies formed and evolved from the earliest epochs to the present day. Surveys have revealed the strong growth in the cosmic star formation rate density from  $\mathrm{z \sim 0-1}$, with a peak in stellar mass build-up at $\mathrm{z \sim 1-3}$, also known as ``cosmic noon". This epoch is also when the cosmic black hole accretion rate peaks \citep[for a review see][]{madau2014}. The present-day Hubble sequence appears to be the result of multiple physical processes that build-up, quench and transform galaxies. These processes depend on mass and redshift and operate on different timescales, and in different environments.  \\
While we have made a lot of progress, we do not yet know how all of these pieces fit together.  In particular, studying galaxy evolution in the context of the cosmic  web requires surveys deep enough to reach below $\mathrm{M^*}$ at cosmic noon, yet covering $\mathrm{>10\, deg^2}$ for a representative volume allowing for the study of galaxies in different environments as well as the study of rare populations \citep[see e.g.][]{Krefting2020}. {\it Spitzer} IRAC coverage is critical in allowing for more accurate stellar population parameters \citep[e.g.,][]{muzzin2009}. For example, for intermediate to high-redshift galaxies ($0.8<z<3$), IRAC imaging are absolutely critical for the UVJ diagnostic \citep[e.g.][]{whitaker2013}, as the rest-frame V - J color cannot be directly measured.  \\
IRAC photometry in combination with precise photometric redshifts was essential to prove galaxy bi-modality and the existence of quiescent galaxies out to $z\sim 3.5$ \citep[e.g.,][]{labbe2005,feldmann2017, hill2017,caputi2017,forrest2018,merlin2018,sherman2021}. Moreover, IRAC imaging has been the only way, until the recent advent of the James Webb Space Telescope, to probe the rest-frame optical wavelengths of very high redshift galaxies \citep[up to $z\sim 10$ , e.g.,][]{labbe2010,labbe2013,strait2020,laporte2021, stefanon2021, bouwens2022, stefanon2022b}. 
Figure\,\ref{f:area_depth} shows a comparison of {\it Spitzer} surveys covering the full area vs. depth parameter space. {\it Spitzer} surveys that balance the requirements of wide ($>$10deg$^2$) area with sufficient depth ($>$23\,mag or 2\,mJy) needed to reach below $M^*$ at cosmic noon and $z\sim5-6$ for massive galaxies include SERVS \citep{mauduit2012}, as well as the {\it Spitzer} coverage of the planned Vera Rubin Observatory survey DeepDrill fields \citep{Lacy2021}, and the Euclid Deep Field South \citep{Laureijs2011,Scarlata2019}. For galaxy evolution studies and for the derivation of robust stellar population properties, multi-wavelength coverage from the $u$-band to the optical need to complement the IRAC data. \\
Until the advent of the next generation facilities like the Vera Rubin Observatory, the HSC-Deep survey \citep{aihara2019} provides the highest quality $grizy$ coverage down to $r$ of 27.4 AB mag across 27$\,$deg$^2$ split among 4 fields: XMM-LSS, E-COSMOS, Deep2-F3, and EN1\footnote{For clarity, the expanded names of these fields are: the XMM Large Scale Structure field (XMM-LSS), the Extended COSMOS field (E-COSMOS), the Deep2 survey Field 3 (Deep2-F3) and the ELAIS survey North1 field (EN1).}. This is a three tiered imaging survey that uses the Hyper Suprime-Cam \citep[HSC;][]{miyazaki1998} designed to address a wide range of astrophysical questions. These four fields also have matching depth $U$-band coverage from the CLAUDS survey \citep{sawicki2019}. However, their IRAC coverage was more patchy. All of XMM-LSS as well as part of EN1 already have sufficiently deep IRAC coverage through SERVS and {\sl Spitzer} DeepDrill \citep{mauduit2012,Lacy2021} as does the central part of the E-COSMOS field \citep{scoville2007}. \\
In this paper, we present additional {\sl Spitzer} observations that complete the IRAC coverage of the HSC-Deep fields to an area of 17~deg$^2$. We also present and publicly release scientific mosaics, coverage maps and basic SExtractor photometry thereof. We combine our new data with re-processed archival data in these fields. The paper is organized as follows. Section \ref{s:obs} describes all the programs that have been used to create single, contiguous deep images in the 3.6 and 4.5$\mu m$ bands. Section\,\ref{s:mosaics}, describes how the mosaics and coverage maps were produced as well as discusses the image quality and depth. Section\,\ref{s:photometry} describes the SExtractor photometry, and the associated number counts. Section\,\ref{s:discussion} presents a discussion of the projected scientific impact of these data.  All coordinates refer to the J2000 system. All magnitudes are in the AB system.

\section{Observations}\label{s:obs}
\subsection{Additional {\sl Spitzer} IRAC observations}\label{ss: new_irac}
In Cycle 14, we were awarded 488.33 hours on the {\sl Spitzer} Space Telescope to extend the {\sl Spitzer} IRAC coverage in E-COSMOS, Deep2-F3 and EN1 (PID:14081; PI: A. Sajina).  The observing strategy followed that of the SERVS  \citep{mauduit2012} and {\sl Spitzer} DeepDrill surveys \citep{Lacy2021}, with six small cycling dithers per pointing, using 100s frametime. These \textit{Astronomical Observation Requests} or AORs tile the desired areas with small overlaps to minimize gaps given the potential differences in relative orientation. The AORs are grouped into two observing epochs, separated by $\approx$ 6 months, which helps remove any moving objects such as asteroids. In addition, the observations of the two channels are carried out simultaneously but are slightly offset on the sky. Every six months, the orientation of the telescope rotates by 180$^{\circ}$, effectively flipping the relative orientation of the two channels. Thus the two epochs help maximize the area of overlap between the two channels. A summary of the new observations in these fields is presented in Table\,\ref{tab:observations}. This summary includes the number of AORs per field and per epoch.

\begin{deluxetable}{cccc}
	\tablewidth{0pt}
	\tablecaption{Summary of new observations
		\label{tab:observations}}
	\tablehead{\colhead{Field/epoch} & \colhead{Obs. date}  & \colhead{\# AORs} & \colhead{Integration time [s]}}
	\startdata 
	E-COSMOS: &  &    \\
	Epoch 1 &  2018/09/13-2018/09/20  & 48 &  600\tablenotemark{a} \\
	Epoch 2 & 2019/03/16-2018/03/30   & 48 &  600 \\
	\hline
	DEEP2-F3:&  &  \\
	Epoch 1 & 2018/09/28-2018/10/13  & 73 & 600 \\
	Epoch 2 & 2018/03/11-2018/03/19  & 72 & 500\tablenotemark{b} \\
	\hline
	EN1:& & \\
	Epoch 1 & 2018/11/30-2018/12/06  & 38  & 500 \\
	Epoch 2 & 2019/06/11-2019/06/19  &  38 & 400  \\
	\hline
	\enddata
\tablenotetext{a}{This is the per pixel exposure time in both channel1 and channel2.}
\tablenotetext{b}{Fields with existing shallower coverage, including SWIRE in EN1 and SpIES in Deep2-F3, have slightly lower exposure times.}
\end{deluxetable}

\subsection{Archival IRAC data \label{ss:archival_irac}}

In this section,  we discuss the archival {\sl Spitzer} IRAC data that have been combined with our new observations to produce our final IRAC mosaics. 

\begin{figure*}[h!]
\centering
\includegraphics[scale=0.57,clip=true]{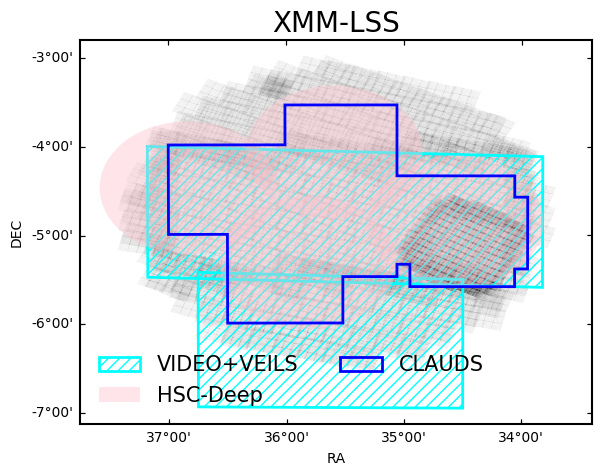}
\includegraphics[scale=0.57,clip=true]{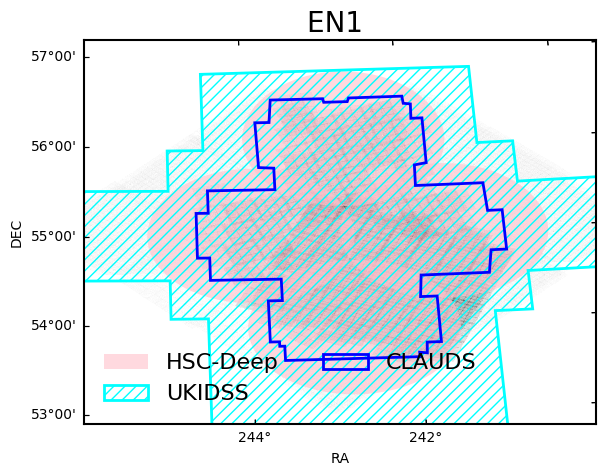}\\
\includegraphics[scale=0.57,clip=true]{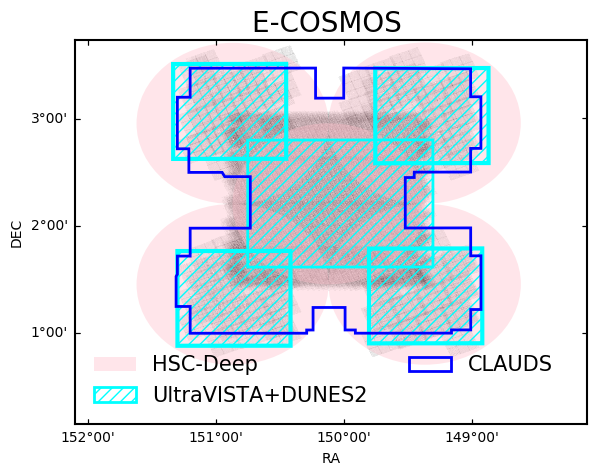}
\includegraphics[scale=0.57,clip=true]{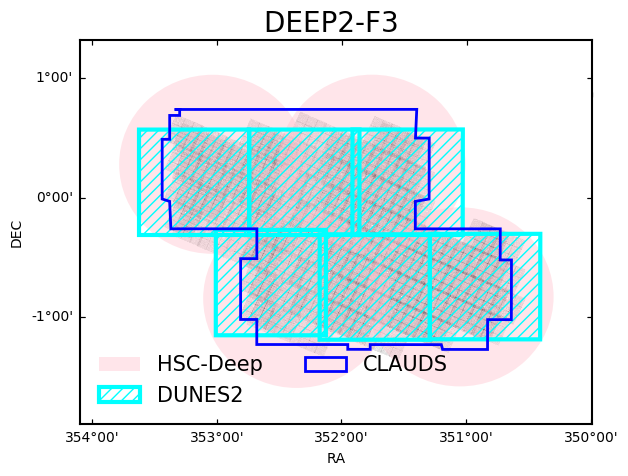}
\caption{The key ancillary data for the four HSC-Deep fields. Greyscale background shows the IRAC exposure maps, with overlaid the footprints of CLAUDS $U$-band (blue outline), HSC-Deep {\it grizy} (pink), and several near-IR surveys (hashed cyan). The total area of overlap of CLAUDS+HSC-Deep+near-IR+IRAC is $\approx$18\,deg$^2$. 
\label{fig:footprints}}  
\end{figure*}

\begin{itemize}
\item DEEP2-FIELD3 (DEEP2-F3) is a $\mathrm{4 \, deg^2}$ field at RA=23h, DEC=-00d. This field  was previously mapped by SpIES \citep{timlin2016} which extended over about 100\,deg$^2$ of the SDSS Stripe82 field (PID 90045; PI G. Richards). These data however are shallow with only 60s exposures per pointing. These data are so shallow that we decided to not combine them with our observations.
\item E-COSMOS is a $\mathrm{5 \, deg^2}$ field 
at RA=10h, DEC=2d whose central 2\,deg$^2$ covers the original COSMOS field \citep{scoville2007}. COSMOS has previous IRAC data coverage coming from multiple surveys, e.g. S-COSMOS \citep{sanders2007}, the Spitzer Extended Deep Survey \citep[SEDS]{ashby2013}, S-CANDELS, \citep{ashby2015}, Star Formation at 4 $<$ z $<$ 6 from the Spitzer Large Area Survey with Hyper-Suprime-Cam \citep[SPLASH;][]{steinhardt2014}, and Spitzer Matching survey of the UltraVISTA ultra-deep Stripes \citep[SMUVS;][]{ashby2018}.

\item ELAIS (European Large Area ISO Survey) North 1 (EN1) is a $\sim$ 8 $deg^2$ field at RA=16h DEC=+55d that was previously imaged in its entirely in both 3.6 and 4.5$\mathrm{\mu m}$ as part of the Spitzer Wide-area InfraRed Extragalactic survey \citep[SWIRE;][]{lonsdale2003}. However these data are shallow with only 80s exposures per pointing. A subsection of 2\,deg$^2$ of this field has exposures of 1200s per pixel from the Spitzer Extragalactic Representative Volume Survey \citep[SERVS;][]{mauduit2012}. Our combined mosaic for EN1 includes both SERVS and SWIRE in addition to the new observations described in Table\,\ref{tab:observations}. The SERVS data in EN1 were obtained early in the {\sl Spitzer} warm mission and subsequently the absolute calibration of the instrument was adjusted. To correct for this calibration difference, multiplicative factors of 1.041 in the $3.6\mu$m channel and 0.980 in the $4.5\mu$m channel were applied to the SERVS data before including them in the combined mosaics as described in the following sections \citep[see also][]{Lacy2021}. 
\end{itemize}
In Table \ref{tab:prog}, we list all the programs that have been used to derive the master mosaics in COSMOS and EN1.

\begin{deluxetable}{lcccc}
	\tablewidth{0pt}
	\tablecaption{Archival IRAC data
		\label{tab:prog}}
	\tablehead{\colhead{Field} & \colhead{Name} & \colhead{PID}  & \colhead{\# AORs} & \colhead{Reference}}
	\startdata 
	E-COSMOS: & & & &\\
	& S-CANDELS & 80057  & 100 & \citealt{ashby2015} \\  
    & S-COSMOS & 20070 & 52 & \citealt{sanders2007}\\
    & SEDS & 61043 & 36 & \citealt{ashby2013} \\
    & SMUVS & 11016 & 378 & \citealt{ashby2018}\\
    & SPLASH & 90042 & 563 & \citealt{steinhardt2014} \\
    & COMPLETE & 13094 & 356 & \citealt{labbe2016}\\
    & COMPLETE2 & 14045 & 34 & \citealt{stefanon2018}\\
	\hline
	EN1:& & & &\\
	& SWIRE & 80096 & 44 & \citealt{lonsdale2003} \\
    & SERVS & 61050 & 91 & \citealt{mauduit2012}\\
	\hline
	\enddata
\end{deluxetable}

\begin{figure}
    \centering
    \includegraphics[scale=0.45]{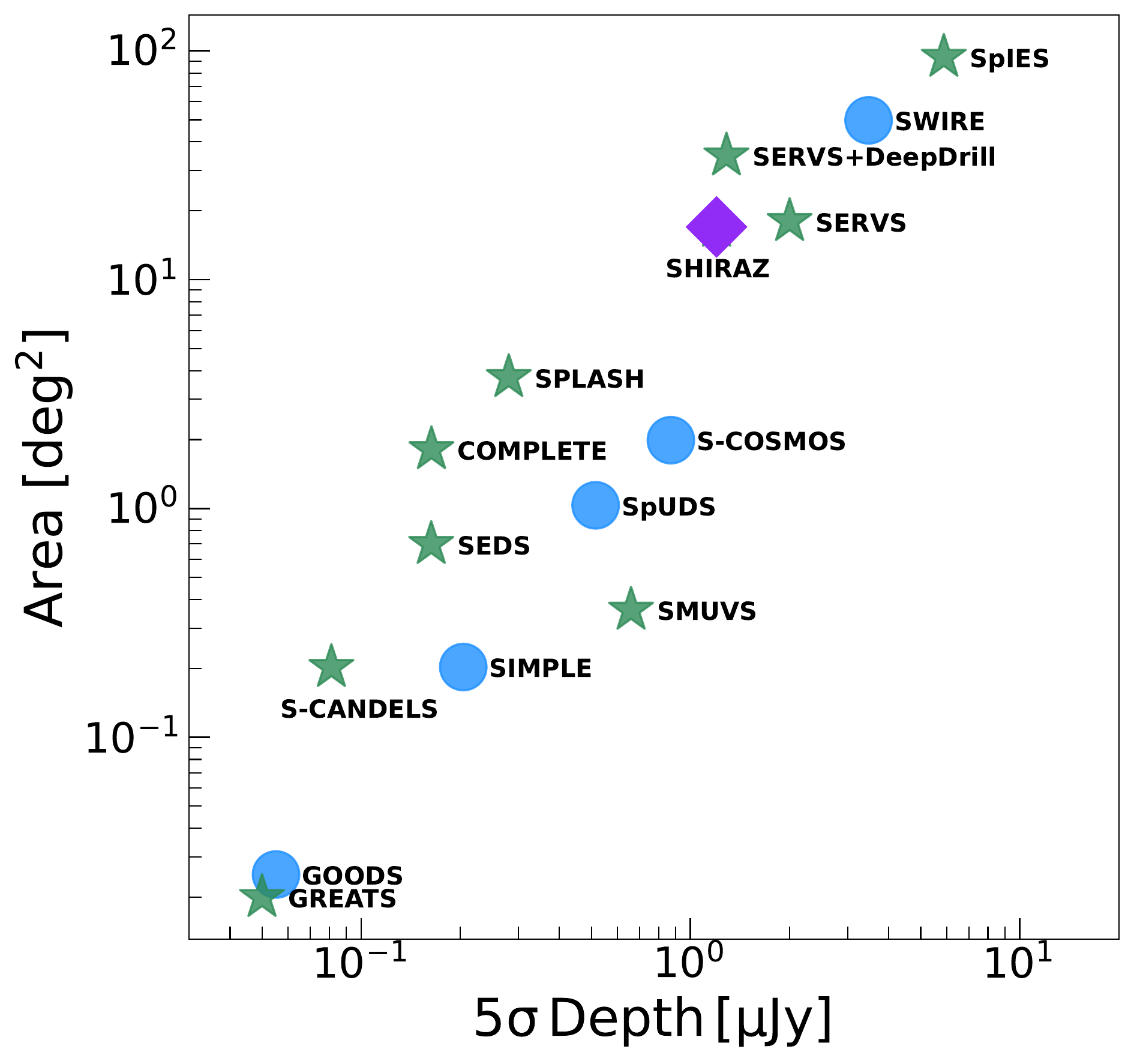}
    \caption{Depth versus area for extragalactic surveys at 3.6$\mu$m. Stars indicate surveys taken during the post-cryogenic phase of Spitzer as Exploration Science or Frontier Legacy surveys. The circles show surveys taken during the cryogenic mission of Spitzer as cyan circles. Our program SHIRAZ is marked as a magenta diamond. 
    References: SpIES \citep{timlin2016}, SWIRE \citep{lonsdale2003}, SERVS-DeepDrill \citep{Lacy2021},
    SERVS \citep{mauduit2012},SPLASH \citep{steinhardt2014}, S-COSMOS \citep{sanders2007}, COMPLETE \citep{labbe2016}, SpUDS \citep{kim2011}, SEDS \citep{ashby2013}, SMUVS \citep{ashby2018}, S-CANDELS \cite{ashby2015}, SIMPLE \citep{damen2011}, GOODS \citep{Dickinson+03}.  }
    \label{f:area_depth}
\end{figure}

Figure\,\ref{fig:footprints} shows the IRAC coverage maps of all four HSC-Deep fields. For completeness, this includes XMM-LSS, which was observed as part of the {\sl Spitzer} DeepDrill survey \citep{Lacy2021}. The coverage maps for the other three fields are discussed in more detail in the subsequent sections of this paper. For all fields, we overlay the most crucial ancilary data including the HSC-Deep, CLAUDS and near-IR surveys. The later include the VIDEO survey \citep{jarvis2013}, the VEILS survey and the UKIDSS survey \citep{Lawrence2007}. It can be seen that with the addition of our new IRAC observations, we now have deep IRAC coverage to go with the areas of overlapping $U$ through near-IR coverage. Note that in this paper we only present the IRAC mosaics for EN1, Deep2-F3, and E-COSMOS. The mosaic for XMM-LSS is presented in \citet{Lacy2021}.  
Figure~\ref{f:area_depth}, shows the area and sensitivity of our mosaics (see Section~\ref{ss_depth}) in the context of other surveys at $3.6 \mu m$. 

\section{Mosaic construction and characterization}\label{s:mosaics}
\subsection{Data reduction}\label{ss:reduction}
We downloaded the {\sl Spitzer} pipeline processed data for all AORs (see Table\,\ref{tab:observations}) observed within this program as well as from archival programs in these fields from the Spitzer Heritage Archive (SHA). All AORs observations are reduced to basic calibrated data (cBCD), using the Spitzer Science Center (SSC) calibration pipeline. In this step, the SSC subtracts dark current, performs
detector linearization and flat-fielding, corrects for artifacts like column pull-ups and downs, multiplexer bleed, first frame effect etc., Then it provides  uncertainty estimates, bad pixel mask, and cosmic-ray rejection masks for each frame. Moreover, the images are absolutely calibrated into physical units (i.e., $MJy/sr =  10^{-17} erg s^{-1} cm^{-2} Hz^{-1} sr^{-1}$). \\
Then, we use the reduction pipeline described in \citet{labbe2015}, \citet{annunziatella2018} and \citet{stefanon2021} to post-processed the cBCDs. Basic calibrated data from different programs were generated using different SSC pipeline versions over the years. This particularly affects astrometry, image distortion refinements, and artifact correction. By post-processing all cBCDs with the same reduction pipeline, we overcame these issues in our final mosaics.
The reduction process consists in a two- pass procedure, where each AOR is reduced independently. The first pass includes: initial background and bias structure estimation from a median of all the frames in the AOR and the consequent subtraction; the correction of column pull-up and pull-down introduced by cosmic rays and bright stars; persistence masking and muxbleed correction rejecting all highly exposed pixels in the subsequent four frames. \\
The second pass includes cosmic-ray rejection, astrometric calibration and an accurate large-scale background removal. Cosmic rays are removed by using an iterative sigma clipping method. The background level is first estimated as the median of the frames in each AOR masking sources and outlier pixels. Then, it is refined  by clipping pixels associated to objects and subtracting the mode of the background pixels.
Single background-subtracted frames are then combined into a mosaic (using WCS astrometry to align the images). In this step, we masked bad pixels and applied a distortion correction in WCS. This image, in combination with the reference image, is then used to refine the pointing of the individual mosaics. The individual pointing-refined frames are registered to and projected on the reference image. The reference images for the three fields are the $z-$band images from the HSC-Deep survey from the second public release (pdr2). For these images, the astrometric calibration was carried out against the Pan-STARRS1 DR1 catalog \citep{aihara2019, chambers2016}.  
The pdr2 website does not release the entire mosaic of each field, but each of them has been divided in tracts of 1.7 $\times$ 1.7 $\mathrm{deg^2}$ \citep{aihara2019}, and each tract in turn into 81 patches of $\mathrm{\sim 12 \times 12}$ arcmin. We downloaded the patches related to each field from the website and combined them in tracts using the python code available on the website. Then, we combined all the tracts of a field using SWarp \citep{Bertin2002}, while resampling them to a pixel scale of 0.6$\arcsec$/pixel (0.5$\times$ the IRAC original pixel scale).
For each field, we produce different end-type images: 
\begin{itemize}
    \item Single epoch mosaics;
    \item Epoch 1 and Epoch 2 combined mosaic;
    \item Combined data of all the observations available in that field. 
\end{itemize}
While we made a master mosaic of E-COSMOS that includes all of IRAC data, most of our analysis is based on the combination of S-COSMOS plus our new observations which pad-out the `corners' of the classical COSMOS field. We chose to do so in order to keep to a relatively uniform depth. 

\begin{figure*}[ht]
	\centering
	\includegraphics[width=\textwidth]{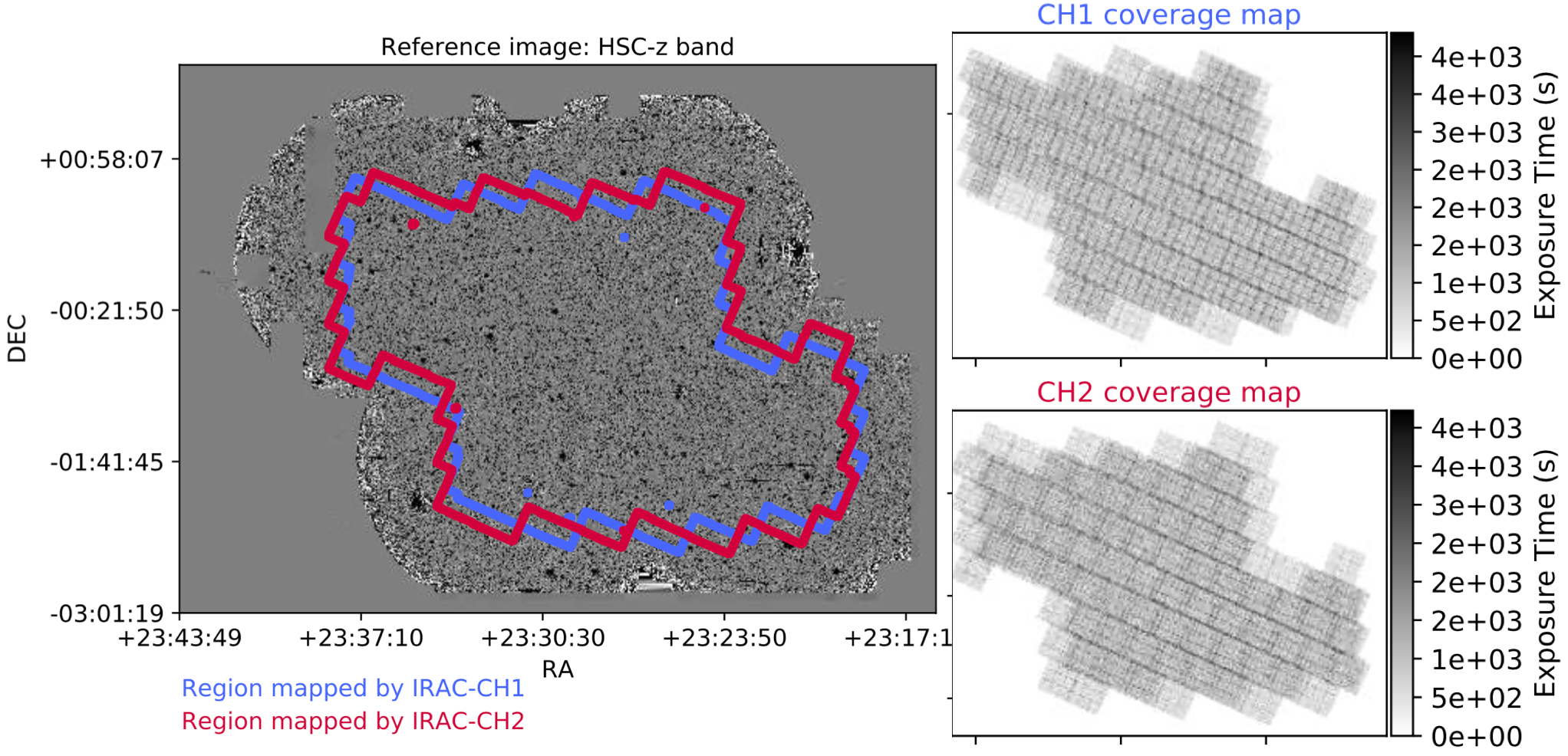}
	\caption{The left-hand side shows the reference z-band mosaic for the DEEP2-F3 field overlaid with the 3.6$\mu m$ and 4.5$\mu m$ mosaic contours. This shows the good degree of overlap between the mosaics of both channels thanks to the two observing epochs. The right-hand side shows the coverage maps for the DEEP2-F3 field for both channels. These only include our new {\sl Spitzer} observations and therefore help illustrate our observing strategy. These clearly show the tiling of the individual AORs each of which consists of 3$\times$3 IRAC pointings. }
	\label{f:DEEP2-F3} 
\end{figure*}

\begin{figure*}[ht]
	\centering
	\includegraphics[width=\textwidth]{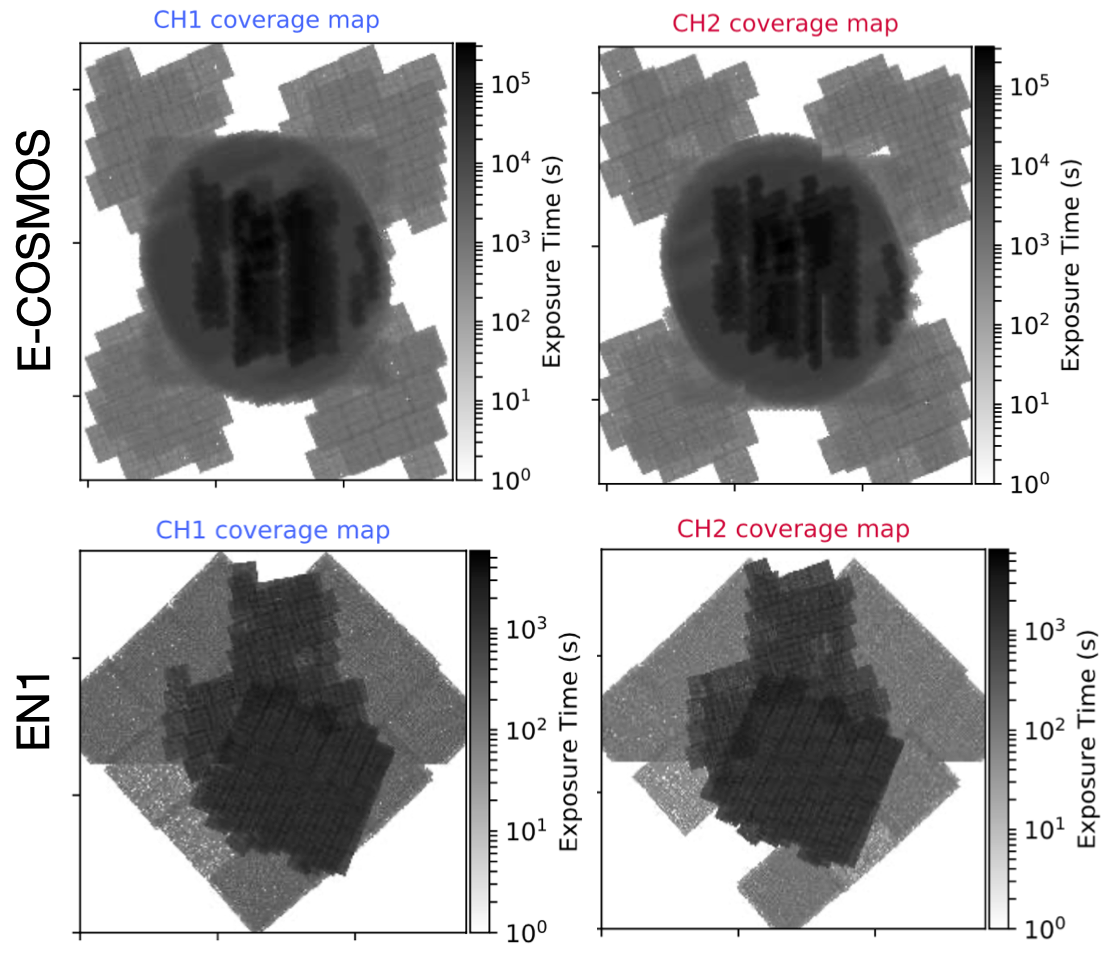}
	\caption{The coverage maps for E-COSMOS (top) and EN1 (bottom). Here we have incorporated both our new as well as archival IRAC data leading to a inhomogeneous depth. For example the shallower area on the outside of the EN1 mosaic comes from the SWIRE survey, but the irregular shaped deeper patches are all within the HSC-Deep footprint (see Figure\,\ref{fig:footprints}). These E-COSMOS coverage maps include all available IRAC data, although to avoid the large contrasts in depths in this field, we did our image quality assessments on the new data plus S-COSMOS alone (which is what is shown in Figure\,\ref{fig:footprints} for E-COSMOS).}
	\label{f:ecosmos_en1_cov}
\end{figure*}

The left-hand side of Figure\,\ref{f:DEEP2-F3} shows for illustration the final IRAC EP1+EP2 mosaics relative to the reference HSC z-band image.  This figure illustrates the good overlap between the channel1 and channel2 coverage of each field thanks to the AORs being split into 2 epochs 6 months apart. The right-hand side of Figure\,\ref{f:DEEP2-F3} shows the coverage maps in this field based only on our new data. These both illustrate the observing strategy as well as the highly uniform coverage in DEEP2-F3 where the mosaic is essentially all our own data. By contrast, Figure\,\ref{f:ecosmos_en1_cov} shows the more variable exposure in E-COSMOS (especially the deeper data in the central 2\,deg$^2$) and EN1 (where the shallower outskirts are the SWIRE data, whereas the deeper irregular shaped center is SERVS+our data).

\subsection{Mosaic quality verification}\label{ss:prop}
In this section, we describe the quality of the final mosaics in both channels for each of the three fields presented in the previous section. To this end, we use all of the available data in EN1 and DEEP2-F3, and only the data corresponding to this program and to S-COSMOS for the E-COSMOS field.

\subsubsection{Point Spread Function}\label{sss:psf}
\begin{figure*}
    \centering
    \includegraphics[scale=0.65]{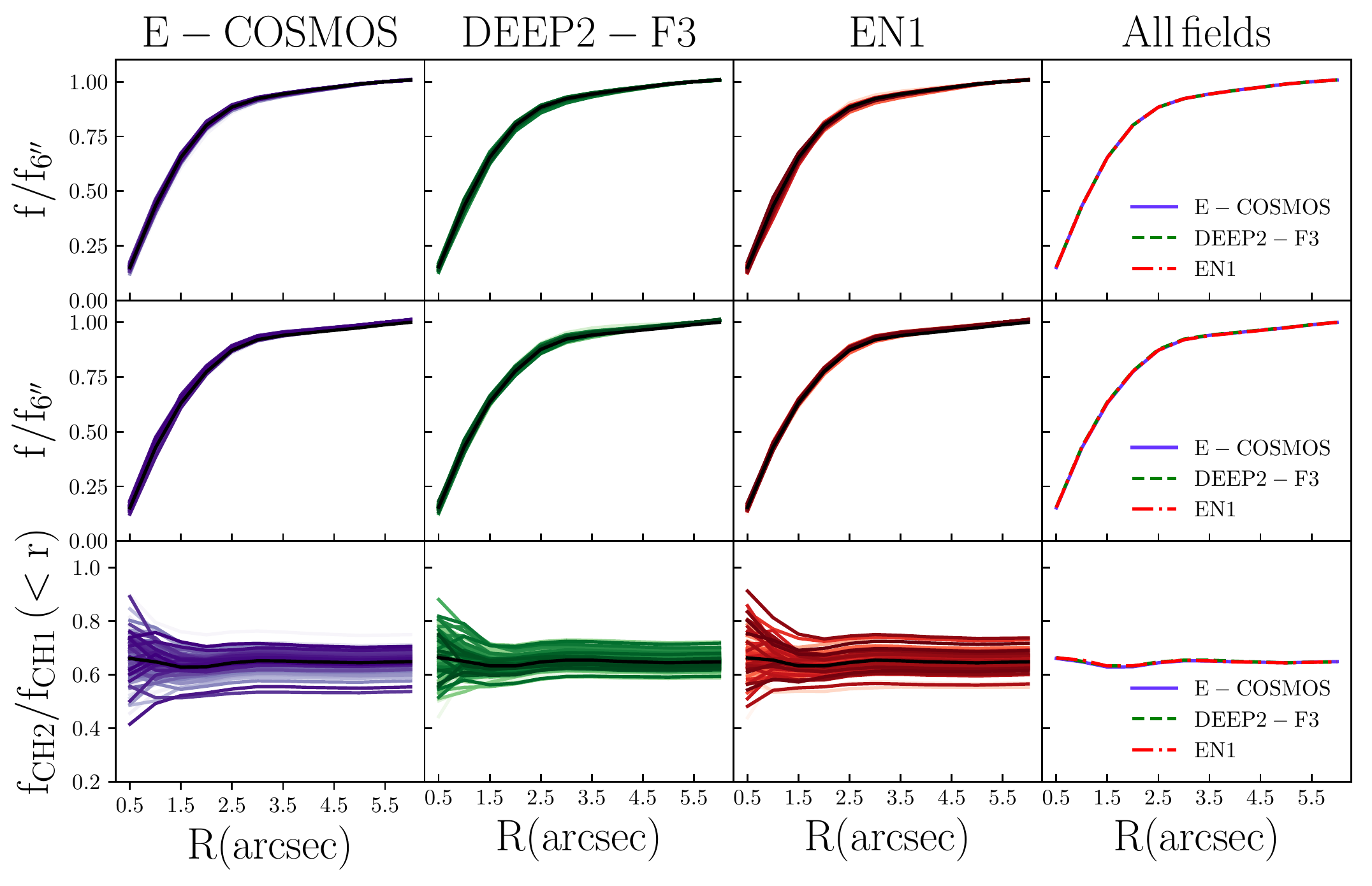}
    \caption{Top row: Growth curves for a sample of $\mathrm{\geq}$ 100 bright unsaturated stars for all three fields at 3.6 $\mu$m normalized to the flux of the star in an aperture of 6\arcsec  diameter. The black curve in each panel represents the median of the distribution. The last panel shows the median growth curve of each field at 3.6 $\mu$m. Middle row: Same as top row but at 4.5 $\mu$m. Bottom row: CH2/CH1 growth curves in the three fields}
    \label{f:psf_visual}
\end{figure*}

To characterize the full-width-half-maximum (FWHM) of the Point Spread Function (PSF) of these mosaics we define a sample  of $\sim$ 100 bright, isolated and unsaturated point sources in each channel and each field. We identified the stars from the SExtractor catalogs in each filter and each field, by combining several methods described in \cite{annunziatella2013} and by visually verifying them.
This sample is used to derive the FWHMs of the IRAC images. The derived FWHM are listed in Table\,\ref{t:psf}. The FWHM varies from 1.69$\arcsec$ in EN1 to 1.77 $\arcsec$ in DEEP2-F3 at 3.6$\mu$m, while it varies from 1.60$\arcsec$ in EN1 to 1.67$\arcsec$ at 4.5$\mu$m. We also used this sample to construct stacked PSF images in each field and each channel. These images showed a fairly stable PSF across the fields. 
For a more quantitative assessment, Figure\,\ref{f:psf_visual} shows the curves-of-growth across each field (and for all fields combined) where the top panels are for the 3.6\,$\mu$m channel whereas the bottom panels are for the 4.5\,$\mu$m channel. Each curve is normalized to the flux in a $\mathrm{6 \arcsec}$ aperture. Such curves-of-growth were constructed for each of our $\sim$100 unsaturated bright point sources. Figure\,\ref{f:psf_visual} also shows relatively little spread among our sample of bright unsaturated stars, consistent with a stable PSF across the mosaics.  We use the median growth curve in each field to derive the median half-light and 75\%-light radii, i.e., the radii that contain 50\% and 75\% of the light, respectively. These values are reported in Table~\ref{t:psf}, together with their $1\sigma$ uncertainties. We checked for possible variations of the growth curves when using EP1, EP2, EP1+EP2, or combined mosaics, and we didn't find any. For this reason, we only report the result obtained when using the combined mosaics (e.g.: EP1+EP2 for DEEP2-F3, EP1+EP2+S-COSMOS for E-COSMOS, and  EP1+EP2+SWIRE+SERVS for EN1). 

\begin{deluxetable}{cccc}
	\tablewidth{0pt}
	\tablecaption{Characteristics of the IRAC observations
		\label{t:psf}}
	\tablehead{\colhead{Filter} & \colhead{FWHM} & \colhead{50\% light radius}  & \colhead{75\% light radius}  \\
		($\mu$m)      & (\arcsec)      & ( \arcsec)   & ( \arcsec) }
	\startdata 
	E-COSMOS:& & &   \\
	3.6 &  $\mathrm{1.72^{+0.14}_{-0.10}}$ & $\mathrm{1.15^{+0.04}_{-0.04}}$ & $\mathrm{1.80^{+0.07}_{-0.06}}$  \\
	4.5 &  $\mathrm{1.61^{+0.10}_{-0.08}}$ & $\mathrm{1.17^{+0.04}_{-0.04}}$ & $\mathrm{1.91^{+0.03}_{-0.04}}$ \\
	\hline
	DEEP2-F3:& & & \\
	3.6 &  $\mathrm{1.77^{+0.14}_{-0.08}}$ & $\mathrm{1.15^{+0.05}_{-0.04}}$  & $\mathrm{1.80^{+0.05}_{-0.04}}$ \\
	4.5 &  $\mathrm{1.67^{+0.12}_{-0.11}}$ & $\mathrm{1.15^{+0.03}_{-0.05}}$  & $\mathrm{1.90^{+0.04}_{-0.07}}$  \\
	\hline
	EN1:& & & \\
	3.6 & $\mathrm{1.69^{+0.22}_{-0.18}}$ & $\mathrm{1.21^{+0.05}_{-0.06}}$ & $\mathrm{1.86^{+0.04}_{-0.05}}$\\
	4.5 & $\mathrm{1.60^{+0.12}_{-0.10}}$ & $\mathrm{1.18^{+0.07}_{-0.07}}$ & $\mathrm{1.93^{+0.03}_{-0.04}}$\\
	\hline
	\enddata
\end{deluxetable}

\subsubsection{Depth}\label{ss_depth}
\noindent To estimate the depth of each mosaic, we use the ``empty aperture'' method to empirically determine the noise properties of our IRAC mosaics, following the same approach as in \citet{annunziatella2018}.  
Briefly, we randomly place a large number of apertures on the noise-normalized images (obtained by multiplying the images by the square root of the coverage maps), reject all apertures falling on sources (identified by SExtractor, see Sect.~\ref{s:photometry}), and measure the flux in the remaining ones. We chose an aperture size of $3\arcsec$ and measure the flux in $\sim$ 4000 apertures in each field and each channel. A histogram of the measured fluxes is constructed for each field  in both $\mathrm{3.6\, \mu m}$ and $\mathrm{4.5\, \mu m}$. These histograms are all well fitted by Gaussians.  The best-fit widths ($\sigma_{best-fit}$) of these histograms can be converted to $\mathrm{3\sigma}$ magnitude depths derived using the empty aperture method according to the Equation:
\begin{equation}
\mathrm{depth(3\sigma)[AB] = -2.5 \log{\Big(\frac{3\sigma_{best-fit}}{\sqrt{\mathrm{COVERAGE\, MAP}}}\Big)} + ZP,}
\label{eqn:depth}
\end{equation}
where ZP is 20.04 for all images. Note that here the division by the square root of the coverage map (aka the exposure time per pixel) reverses the procedure above when we generated noise-normalized images.  

\begin{figure*}[ht]
\centering
	\includegraphics[scale=0.2]{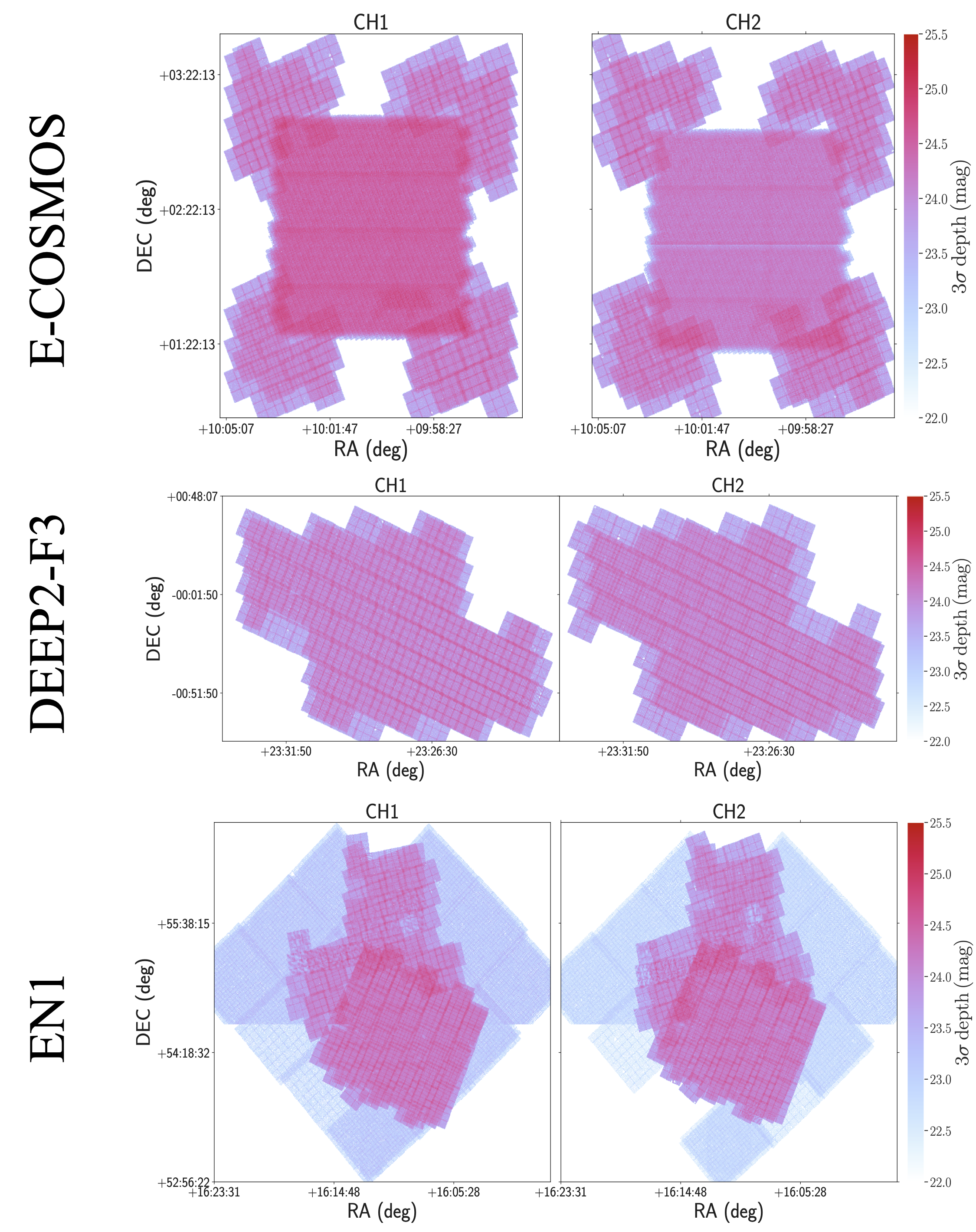}
	\caption{3$\sigma$ magnitude depths of the $\mathrm{3.6\, \mu m}$ (left) and $\mathrm{4.5\, \mu m}$ (right) mosaics. These are computed from Equation\,\ref{eqn:depth} based on the width of the distribution of fluxes measured in $\sim$4000 empty apertures from the noise-normalized images (see text for details). We see that in the areas of interest (overlapping with the HSC-Deep layer) our mosaics have a 3\,$\sigma$ depth of $\sim$24 [AB] or better.}
	\label{f:empty_depths}
\end{figure*}

Figure\,\ref{f:empty_depths} shows the 3\,$\sigma$ magnitude depth for each field and each channel. As expected, regions with higher exposure times (see Figures\,\ref{f:DEEP2-F3} and\ref{f:ecosmos_en1_cov}) also have fainter $3\sigma$ magnitude depths.  Table ~\ref{t:ap} lists the 15th, 50th, and 75th percentiles of the 3$\sigma$ magnitude depths in both $\mathrm{3.6\, \mu m}$ and $\mathrm{4.5\, \mu m}$ in an aperture of D=3$\arcsec$ as derived with the empty aperture method. As expected the depth is shallowest for the EN1 field, channel 2, where we have the lowest exposure time (see Table\,\ref{tab:observations}).  We also converted these to the more conventional 5$\sigma$ values and found that on average the 5$\sigma$ depth is 23.7 at 3.6$\mu$m and 23.3 at 4.5$\mu$m. This is comparable to the depths of the SERVS \citep{mauduit2012} and {\sl Spitzer} DeepDrill \citep{Lacy2021} surveys.  

\begin{deluxetable*}{cccc}
	\tablewidth{0pt}
	\tablecaption{Depths of the IRAC observations
		\label{t:ap}}
	\tablehead{\colhead{Filter} & \colhead{15th percentile of 3$\sigma$ depth\tablenotemark{a}}  & \colhead{Median of 3$\sigma$ depth\tablenotemark{b} \tablenotemark{a}} & \colhead{85th percentile of 3$\sigma$ depth\tablenotemark{a}} \\
		($\mu$m)          & ( AB)   & ( AB)  & ( AB)}
	\startdata 
	E-COSMOS: &  &    \\
	3.6 &   23.82 & 24.30 & 24.49  \\
	4.5 &   23.48 & 23.84 & 23.97  \\
	\hline
	DEEP2-F3:&  &  \\
	3.6 &    23.72 & 24.01 & 24.16 \\
	4.5 &    23.39 & 23.74 & 23.89 \\
	\hline
	EN1:& & \\
	3.6 &  22.94 & 23.54 & 24.30\\
	4.5 &  22.44 & 23.14 & 24.05 \\
	\hline
	\enddata
	\tablenotetext{a}{Using a 3$\arcsec$ circular aperture diameter. }
    \tablenotetext{b}{To convert these to 5$\sigma$ simply subtract 0.22 from these numbers. The average 5$\sigma$ depth for the three fields is 23.7 at at 3.6$\mu$m and 23.3 at 4.5$\mu$m. }
\end{deluxetable*}

\section{SExtractor photometry and number counts}\label{s:photometry}
The state-of-the-art approach to extract IRAC photometry from wide and deep surveys consists in the adoption of software that accounts for: 1. PSF spatial variations across the field and 2. source confusion. As described in Section\,\ref{sss:psf}, our mosaics show a fairly stable PSF but with not completely negligible PSF variation across with position. Addressing IRAC source confusion involves a higher resolution ``prior" image that helps define the positions and shapes of the blended sources \citep[e.g.][]{Nyland2017}. We are working on such forced photometry for our mosaics, in the near future and indeed, the resultant photometry will be included in multi-band photometric catalogs to be released from the HSC collaboration. 
For this paper, we perform basic SExtractor \citep{bertin1996} photometry on our mosaics, both the combined ones as well as the single epoch ones. Photometry extracted from the IRAC images alone (as here) without higher angular resolution priors is subject to source confusion. We refer the reader to \citet{Lacy2021} for more detail on how confusion affects their IRAC photometry given that they have data of similar depth.  With this caveat in mind, the basic photometry presented here serves the purpose of further characterizing the mosaics including the resultant number counts in each field. Future catalogs will only consider the combined deeper mosaics.  \\  
First, we ran SExtractor on the single epoch images from this work and compare the photometry between the two different epochs. We found significant (above 3$\sigma$), flux variation for only 0.2\% of sources with $m_{3.6}>18$, some of which may be genuine variable sources and some due to photometry errors. \\
Then, we ran SExtractor on the mosaics in the EN1, E-COSMOS and DEEP2-F3 fields. As in Sect.~\ref{ss:prop}, we consider only data from this program for DEEP2-F3, the combination of this program and S-COSMOS in E-COSMOS and the combination of this program, SWIRE and SERVS for EN1.\\
The parameters used for the catalog extraction are shown in Table~\ref{t:sex} and are the same used in \citealt{Lacy2021}.

\begin{deluxetable}{cc}
	\tablewidth{0pt}
	\tablecaption{Sextractor main parameters. \label{t:sex}}
	\tablehead{\colhead{Parameter} & \colhead{Value}}
	\startdata 
    DETECT\_MINAREA & 5.0\\
    DETECT\_THRESH & 1.0\\
    ANALYSIS\_THRESH & 0.4\\
    DEBLEND\_NTHRESH & 64\\
    DEBLEND\_MINCONT & 0.0005\\
    BACK\_SIZE & 16 \\
    BACK\_FILTERSIZE & 3 \\
    BACKPHOTO\_TYPE & LOCAL\\
	\enddata
\end{deluxetable}

Overall, this analysis resulted in 434K total sources in Deep2-F3, 243K sources in EN1 and 630K sources in E-COSMOS. These numbers are lower limits on the IRAC sources we will eventually have in these fields, as the forced photometry will allow us to reach deeper levels than blank sky photometry such as the present SExtractor run \citep[][]{Nyland2017}.\\

\subsection{Astrometric accuracy}\label{sss:wcs_accuracy}
We matched the obtained catalogs to Gaia Data Release 3 \citep[DR3;][]{torra2021, lindegren2018}. The IRAC pointing is calibrated using the HSC-z band reference image. We matched the position of the sources in our catalogs with those from Gaia DR3 using a 1.0\arcsec match radius. Around $\sim$ 4\% of the sources in the three fields have counterparts in Gaia DR3. The results are shown in Table~\ref{t:wcs}, where we list the median systematic offset between Spitzer and Gaia DR3 positions (R.A.) and (Dec.), along with the scatter $\sigma$ (R.A.) and $\sigma$ (Dec.), representing the positional accuracy of a typical source in our fields.
All systematic offsets are $<$ 0.1\arcsec.

\begin{deluxetable}{ccccc}
	\tablewidth{0pt}
	\tablecaption{Astrometric accuracy
		\label{t:wcs}}
	\tablehead{\colhead{Filter} & \colhead{$\Delta$ (R.A.) }  & \colhead{$\Delta$ (Dec.)} & \colhead{$\sigma(\Delta (R.A.))$} & \colhead{$\sigma(\Delta (Dec.))$} \\
		($\mu$m)  & (\arcsec) & ( \arcsec)  & ( \arcsec) & ( \arcsec)  }
	\startdata 
	E-COSMOS: & & & &   \\
	3.6 &  -0.04 & -0.05 & 0.19 & 0.19 \\
	4.5 &  -0.04 & -0.05 & 0.20 & 0.19 \\
	\hline
	DEEP2-F3:& & & &\\
	3.6 &  0.02 & -0.08 & 0.23 & 0.21 \\
	4.5 &  0.03 & -0.07 & 0.24 & 0.21  \\
	\hline
	EN1: & & & & \\
	3.6 & -0.06 & -0.03 & 0.33 & 0.20\\
	4.5 & -0.03 & -0.03 & 0.35 & 0.20\\
	\hline
	\enddata
\end{deluxetable}

\subsection{Photometric accuracy}\label{sss:phot_accuracy}
\begin{figure*}[h]
    \centering
    \includegraphics[width=\linewidth]{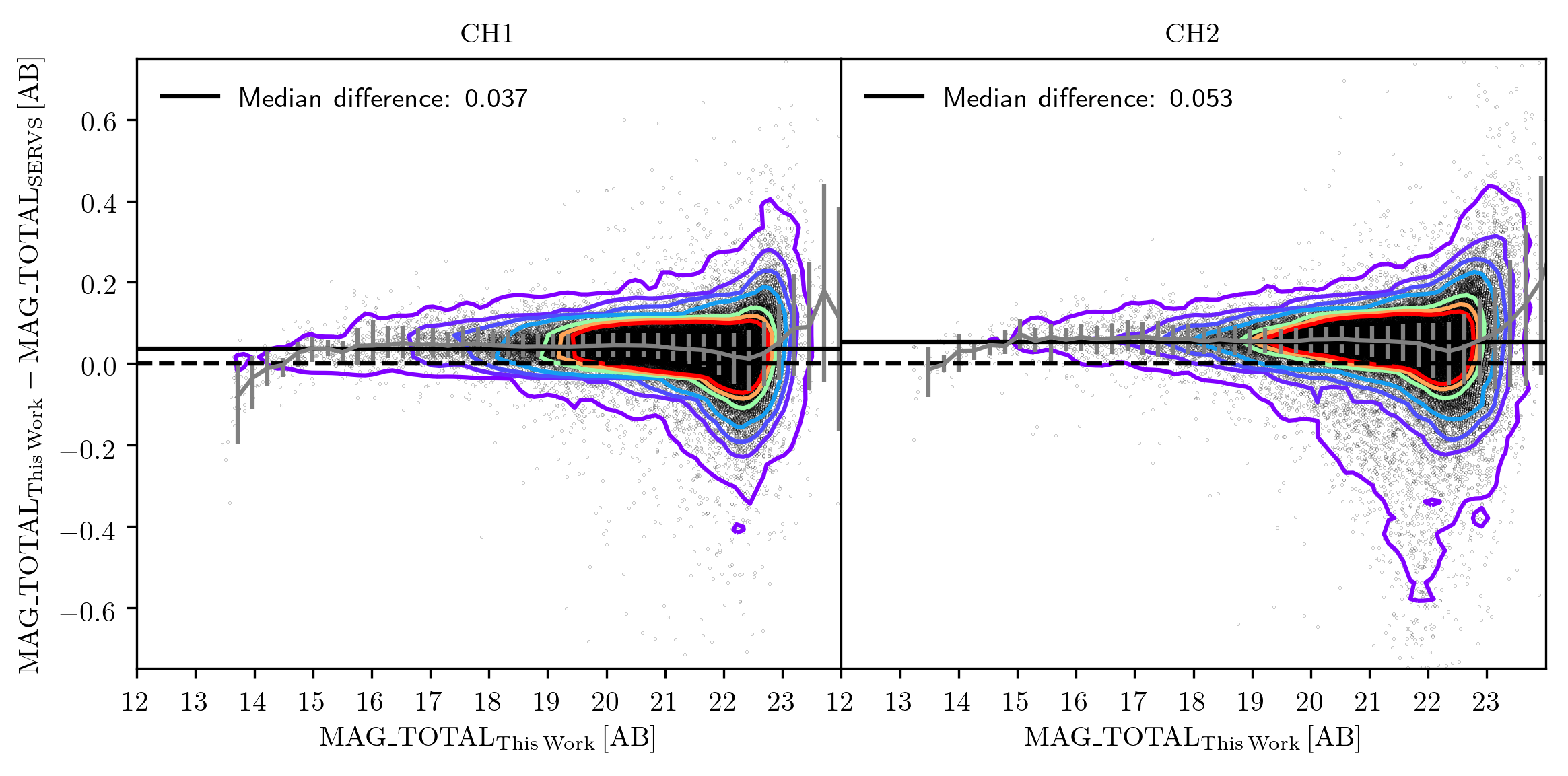}
    \caption{We compare the total magnitudes for objects detected in our new mosaic of EN1 and those from the SERVS catalog \citep{mauduit2012} in the same area in both channels.  Since the SERVS catalog excludes stars \citet{mauduit2012}, in this cross-match stars are also excluded.  Contours, from the inner to the outer, contain 10\%,30\%,50\%,70\%,90\%,95\% and 99\% of the galaxies. 
Overplotted in grey are the running medians and the correspondent $\mathrm{1\sigma}$ confidence interval. We see pretty good agreement with only small median offsets. For channel 1, this is largely explained by the recalibration of the first year of {\sl Spitzer} warm mission data that was not known at the time of the original SERVS catalog construction.}
    \label{f:comparison}
\end{figure*}

Figure~\ref{f:comparison} shows the comparison between the total magnitudes for the objects detected in EN1 and those from the SERVS catalog \citep{mauduit2012} in both the 3.6$\mu$m mosaic (left-hand panel) and the 4.5$\mu$m mosaic (right-hand panel). Our mosaic in EN1 in the area where it overlaps with SERVS is entirely made up of SERVS data (both our mosaic and SERVS co-add the SWIRE data here). Therefore, the only differences are in the data processing and the specific SExtractor settings. 

In order to estimate the total magnitude for the sources in EN1, we multiply the fluxes of the sources in an aperture of $\mathrm{3.9\arcsec}$ diameter for an aperture correction factor derived using the median growth curves shown in the third panel of Fig.~\ref{f:psf_visual}. The total fluxes for the sources in SERVS are provided in the SERVS catalog and are derived using a similar approach as in this work, but with a slightest smaller aperture ($\mathrm{D=3.8\arcsec}$).
While unsurprisingly there are some outliers, the vast majority of the sources show excellent agreement between the earlier SERVS catalog \citep{mauduit2012} and our new SExtractor photometry based on a re-processed EN1 mosaic. 
Figure~\ref{f:comparison} shows this comparison in both channels. The comparison is carried out up to AB=24mag. There are median offsets between the magnitudes which are: 0.04 at 3.6$\mu$m and 0.05 for 4.5$\mu$m. Some of this is due to our accounting for the re-calibration of the 1st year {\sl Spitzer} warm mission data (see Section\,\ref{ss:archival_irac}) which was not known at the time of the original SERVS catalog. This level offsets are also not unexpected given the somewhat different SExtractor settings we use relative to the ones in \citet{mauduit2012}. In the figure, there are overplotted in grey the running medians and the correspondent 1$\sigma$ confidence interval. As it can be seen, there is no trend with magnitude over the entire range.

\subsection{Number counts}\label{ss:counts}

\begin{figure*}
    \centering
    \includegraphics[scale=0.5]{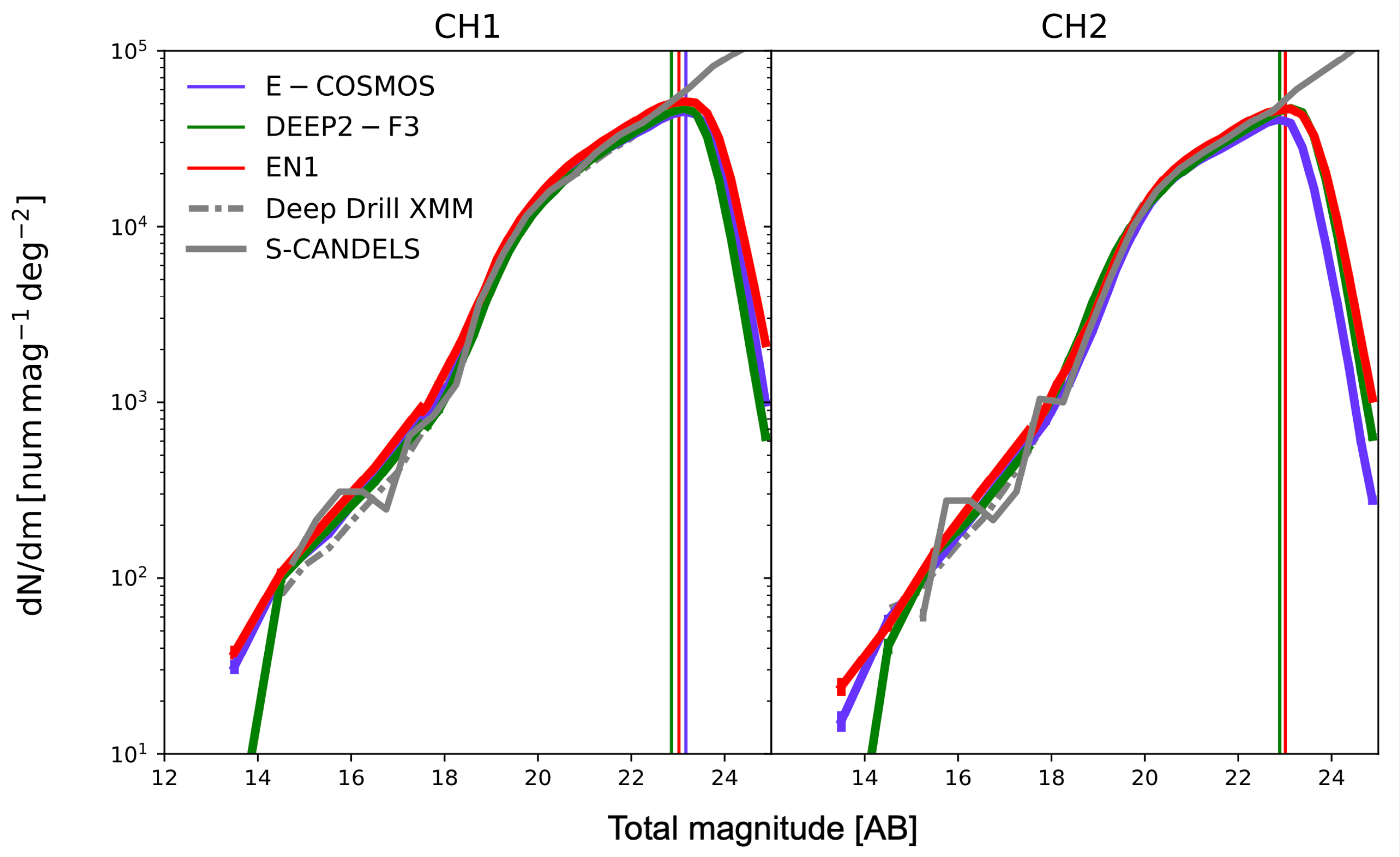}
    \caption{Differential source counts in the E-COSMOS, DEEP2-F3 and EN1 fields measured in the 3.6$\mu$m mosaic (left panel) and in the 4.5$\mu$m mosaic (right panel). The vertical lines are the 50th percentile of the 5$\sigma$ depths. For comparison, we overplot the counts from the S-CANDELS \citep{ashby2015}, and from the Deep Drill \citep{Lacy2021} surveys.}
    \label{f:counts}
\end{figure*}

\noindent Figure~\ref{f:counts} shows the differential source counts in the E-COSMOS, DEEP2-F3 and EN1 fields measured in 3.6$\mu$m and 4.5$\mu$m. The number counts are derived using aperture magnitudes computed within an aperture of diameter 3.9''' scaled to total using the median growth curves shown in Fig.~\ref{f:psf_visual}. The effective area in each field is evaluated by excluding the regions with lower exposure times and shallower depths (i.e. the SWIRE region in EN1, or regions where epochs do not overlap). This leads to an effective area of: 4.41, 2.93 and 3.68 $\mathrm{deg^2}$ for E-COSMOS, DEEP2-F3 and EN1, respectively. We also overplot the values of the 5$\sigma$ depth in each field and each channel scaled to total magnitude using the same approach as above. We compared our differential number counts with those from SERVS and S-CANDELS. In both cases the total magnitudes were obtained by applying a correction function to the fluxes derived in apertures of 3.8'' and 2.4'' diameter respectively for SERVS and S-CANDELS. As we can see from Figure~\ref{f:counts}, the number of detected sources in both channels in EN1 and E-COSMOS is comparable to that of previous surveys, up to the respective 5$\sigma$ limit.

Lastly, we note that we do not remove stars from our counts. The slight upturn seen in the counts at magnitudes $\approx$18 is due to the transition from the regime where stars dominate the counts to where galaxies dominate the counts \citep[see e.g.][]{ashby2015,Lacy2021}.

\section{Discussion}\label{s:discussion}

\subsection{Public Data release}\label{ss:release}
Accompanying this paper, we present the public data release, consisting of reduced images of all IRAC observations in three fields, namely E-COSMOS, DEEP2-F3, and EN1. All data products will be available on the NASA/IPAC Infrared Science Archive (IRSA).
The data release contains the following:
\begin{itemize}
\item Spitzer/IRAC mosaic science images in both 3.6 $\mu$m and 4.5 $\mu$m. The images are astrometrically registered to the reference z-band image from the HSC Subaru Public Data release 2.
\item Coverage maps containing exposure times (seconds) in both 3.6 $\mu$m and 4.5 $\mu$m. 
\item Maps with 3$\sigma$ magnitude depth (as derived in Section~\ref{ss:prop}) at the same WCS and pixel scales of the scientific images.
\end{itemize}

\subsection{Expected Science Impact \label{ss:impact}}
The primary goal for obtaining these new data and constructing these mosaics combining all available data was to provide IRAC 3.6 and 4.5$\mu$m coverage of the HSC-Deep fields. The mosaics presented in this paper represent a total of 17.9\,$deg^2$, which together with the XMM-LSS IRAC coverage through SERVS+Spitzer DeepDrill \citep{Lacy2021} means that the bulk of the 27\,$deg^2$ HSC-Deep fields now has sufficiently deep IRAC coverage to study massive galaxies out to $z\sim6$ as well as reach well below the knee of stellar mass function at cosmic noon. Indeed, adding the IRAC data to the wealth of $U$ through near-IR photometry already obtained in these areas will lead to improved photometric redshifts but even more significant improvements in the derived stellar population parameters such as stellar mass, age and star-formation rate \citep[see e.g.][]{muzzin2009}. This will impact a wide range of galaxy evolution studies to be performed with these data.
The wide area of the HSC-Deep survey makes it well suited for the study of the role of environment in galaxy assembly and quenching since it will sample the full range of environments with high statistical significance. For example, in only 4.5\,$deg^2$ of the XMM-LSS, \citet{Krefting2020} find nearly 400 overdensities consistent with dark matter halos of $M>10^{13.7}M_{\odot}$ including several structures that look like cluster mergers and large filamentary structures. With 5.5$\times$ the area covered in the present data, the potential to explore galaxy evolution in the context of both local and large scale environment is great.  
These data are also well suited to AGN studies including BH-galaxy co-evolution studies. For BH-galaxy co-evolution studies what is critical is that the combined multiwavelength data will allow for both photometric redshifts and stellar population parameters of all potential AGN host galaxies. The IRAC data are critical for accurate estimates of such parameters as described above.  This deep IRAC survey is also potentially useful for finding AGN in dwarf galaxies which may be powered by IMBHs \citep{Satyapal2017}. 

Lastly, the HSC-Deep fields will be the sites of the upcoming PrimeFocusSpectrograph (PFS) galaxy evolution survey \citep{tanaka2017}. The PFS is a highly multiplexed spectrograph with $\approx2500$ fibers being build for the 10m Subaru telescope. The PFS spectra will cover $\lambda\sim0.3-1.3$\,$\mu$m with $R\sim3000$ providing a wealth of information for the targeted galaxies including nebular-line based SFRs, metallicites, AGN content, outflow rates to name a few.  The multi-band photometric coverage of these fields, including IRAC, will allow for the optimal target selection for the PFS as well as providing significant value to the spectroscopic data by allowing for key photometrically-derived parameters such as stellar mass. 

\section{Summary \& Conclusions} 
\begin{itemize}
\item This paper presents IRAC 3.6$\mu$m and 4.5$\mu$m mosaics using both new and archival data for three of the four fields of the HSC-Deep survey (E-COSMOS, DEEP2-F3, and EN1). The mosaics have a median 5$\sigma$ depth of 23.7(23.3) at 3.6(4.5)$\mu$m in AB. The science images, coverage maps and depth maps will all be publicly released concurrently with the publication of this paper.
\item We  perform SExtractor photometry and validate the results both by comparison with earlier catalogs as well as by checking the resultant number counts for consistency with IRAC counts in the literature. 
\end{itemize}

JWST can now provide very deep, high spatial resolution imaging at 3.6 and 4.5 mu using the NIRCAM imager. However, the smaller field of view of NIRCAM, combined with the slow slew rate of the large JWST telescope result in a much slower survey mapping speed than Spitzer/IRAC. Thus, surveys of multiple square degrees of sky in these bands performed by Spitzer will remain the best way to identify large samples of galaxies around cosmic noon  at z$\sim$2. 

\section*{Acknowledgments}

This work has made use of data from the European Space Agency (ESA) mission
{\it Gaia} (\url{https://www.cosmos.esa.int/gaia}), processed by the {\it Gaia}
Data Processing and Analysis Consortium (DPAC,
\url{https://www.cosmos.esa.int/web/gaia/dpac/consortium}). Funding for the DPAC has been provided by national institutions, in particular the institutions
participating in the {\it Gaia} Multilateral Agreement. This research has made use of the NASA/IPAC Extragalactic Database (NED),
which is operated by the Jet Propulsion Laboratory, California Institute of Technology,
under contract with the National Aeronautics and Space Administration. MA acknowledges support by the National Aeronautics and Space Administration (NASA) through an award (RSA 1628138) issued by JPL/Caltech.  AS and DM acknowledge support by NASA under award number 80NSSC21K0630, issued through the Astrophysics Data Analysis Program (ADAP). MA acknowledges financial support from Comunidad de Madrid under Atracci\'on de Talento grant 2020-T2/TIC-19971. XL acknowledges support from NSF grants AST-2108162 and AST-2206499.

\bibliography{bibliography}

\begin{thebibliography}{}
\expandafter\ifx\csname natexlab\endcsname\relax\def\natexlab#1{#1}\fi
\providecommand{\url}[1]{\href{#1}{#1}}
\providecommand{\dodoi}[1]{doi:~\href{http://doi.org/#1}{\nolinkurl{#1}}}
\providecommand{\doeprint}[1]{\href{http://ascl.net/#1}{\nolinkurl{http://ascl.net/#1}}}
\providecommand{\doarXiv}[1]{\href{https://arxiv.org/abs/#1}{\nolinkurl{https://arxiv.org/abs/#1}}}

\bibitem[{{Aihara} {et~al.}(2019){Aihara}, {AlSayyad}, {Ando}, {Armstrong},
  {Bosch}, {Egami}, {Furusawa}, {Furusawa}, {Goulding}, {Harikane}, {Hikage},
  {Ho}, {Hsieh}, {Huang}, {Ikeda}, {Imanishi}, {Ito}, {Iwata}, {Jaelani},
  {Kakuma}, {Kawana}, {Kikuta}, {Kobayashi}, {Koike}, {Komiyama}, {Li},
  {Liang}, {Lin}, {Luo}, {Lupton}, {Lust}, {MacArthur}, {Matsuoka}, {Mineo},
  {Miyatake}, {Miyazaki}, {More}, {Murata}, {Namiki}, {Nishizawa}, {Oguri},
  {Okabe}, {Okamoto}, {Okura}, {Ono}, {Onodera}, {Onoue}, {Osato}, {Ouchi},
  {Shibuya}, {Strauss}, {Sugiyama}, {Suto}, {Takada}, {Takagi}, {Takata},
  {Takita}, {Tanaka}, {Terai}, {Toba}, {Uchiyama}, {Utsumi}, {Wang}, {Wang}, \&
  {Yamada}}]{aihara2019}
{Aihara}, H., {AlSayyad}, Y., {Ando}, M., {et~al.} 2019, \pasj, 71, 114,
  \dodoi{10.1093/pasj/psz103}

\bibitem[{{Annunziatella} {et~al.}(2013){Annunziatella}, {Mercurio}, {Brescia},
  {Cavuoti}, \& {Longo}}]{annunziatella2013}
{Annunziatella}, M., {Mercurio}, A., {Brescia}, M., {Cavuoti}, S., \& {Longo},
  G. 2013, \pasp, 125, 68, \dodoi{10.1086/669333}

\bibitem[{{Annunziatella} {et~al.}(2018){Annunziatella}, {Marchesini},
  {Stefanon}, {Muzzin}, {Lange-Vagle}, {Cybulski}, {Labbe}, {Kado-Fong},
  {Bezanson}, {Brammer}, {Herrera}, {Lundgren}, {Marsan}, {Nonino}, {Rudnick},
  {Saracco}, {Tomer}, {Valdes}, {van der Burg}, {van Dokkum}, {Wake}, \&
  {Whitaker}}]{annunziatella2018}
{Annunziatella}, M., {Marchesini}, D., {Stefanon}, M., {et~al.} 2018, \pasp,
  130, 124501, \dodoi{10.1088/1538-3873/aae796}

\bibitem[{{Ashby} {et~al.}(2013){Ashby}, {Willner}, {Fazio}, {Huang}, {Arendt},
  {Barmby}, {Barro}, {Bell}, {Bouwens}, {Cattaneo}, {Croton}, {Dav{\'e}},
  {Dunlop}, {Egami}, {Faber}, {Finlator}, {Grogin}, {Guhathakurta},
  {Hernquist}, {Hora}, {Illingworth}, {Kashlinsky}, {Koekemoer}, {Koo},
  {Labb{\'e}}, {Li}, {Lin}, {Moseley}, {Nandra}, {Newman}, {Noeske}, {Ouchi},
  {Peth}, {Rigopoulou}, {Robertson}, {Sarajedini}, {Simard}, {Smith}, {Wang},
  {Wechsler}, {Weiner}, {Wilson}, {Wuyts}, {Yamada}, \& {Yan}}]{ashby2013}
{Ashby}, M.~L.~N., {Willner}, S.~P., {Fazio}, G.~G., {et~al.} 2013, \apj, 769,
  80, \dodoi{10.1088/0004-637X/769/1/80}

\bibitem[{{Ashby} {et~al.}(2015){Ashby}, {Willner}, {Fazio}, {Dunlop}, {Egami},
  {Faber}, {Ferguson}, {Grogin}, {Hora}, {Huang}, {Koekemoer}, {Labb{\'e}}, \&
  {Wang}}]{ashby2015}
---. 2015, \apjs, 218, 33, \dodoi{10.1088/0067-0049/218/2/33}

\bibitem[{{Ashby} {et~al.}(2018){Ashby}, {Caputi}, {Cowley}, {Deshmukh},
  {Dunlop}, {Milvang-Jensen}, {Fynbo}, {Muzzin}, {McCracken}, {Le Fevre},
  {Huang}, \& {Zhang}}]{ashby2018}
{Ashby}, M.~L.~N., {Caputi}, K.~I., {Cowley}, W., {et~al.} 2018, ArXiv
  e-prints.
\newblock \doarXiv{1801.02660}

\bibitem[{{Bertin} \& {Arnouts}(1996)}]{bertin1996}
{Bertin}, E., \& {Arnouts}, S. 1996, \aaps, 117, 393,
  \dodoi{10.1051/aas:1996164}

\bibitem[{{Bertin} {et~al.}(2002){Bertin}, {Mellier}, {Radovich}, {Missonnier},
  {Didelon}, \& {Morin}}]{Bertin2002}
{Bertin}, E., {Mellier}, Y., {Radovich}, M., {et~al.} 2002, in Astronomical
  Society of the Pacific Conference Series, Vol. 281, Astronomical Data
  Analysis Software and Systems XI, ed. D.~A. {Bohlender}, D.~{Durand}, \&
  T.~H. {Handley}, 228

\bibitem[{{Bouwens} {et~al.}(2022){Bouwens}, {Illingworth}, {Ellis}, {Oesch},
  {Paulino-Afonso}, {Ribeiro}, \& {Stefanon}}]{bouwens2022}
{Bouwens}, R.~J., {Illingworth}, G., {Ellis}, R.~S., {et~al.} 2022, \apj, 931,
  81, \dodoi{10.3847/1538-4357/ac618c}

\bibitem[{{Caputi} {et~al.}(2017){Caputi}, {Deshmukh}, {Ashby}, {Cowley},
  {Bisigello}, {Fazio}, {Fynbo}, {Le F{\`e}vre}, {Milvang-Jensen}, \&
  {Ilbert}}]{caputi2017}
{Caputi}, K.~I., {Deshmukh}, S., {Ashby}, M.~L.~N., {et~al.} 2017, \apj, 849,
  45, \dodoi{10.3847/1538-4357/aa901e}

\bibitem[{{Chambers} {et~al.}(2016){Chambers}, {Magnier}, {Metcalfe},
  {Flewelling}, {Huber}, {Waters}, {Denneau}, {Draper}, {Farrow}, {Finkbeiner},
  {Holmberg}, {Koppenhoefer}, {Price}, {Rest}, {Saglia}, {Schlafly}, {Smartt},
  {Sweeney}, {Wainscoat}, {Burgett}, {Chastel}, {Grav}, {Heasley}, {Hodapp},
  {Jedicke}, {Kaiser}, {Kudritzki}, {Luppino}, {Lupton}, {Monet}, {Morgan},
  {Onaka}, {Shiao}, {Stubbs}, {Tonry}, {White}, {Ba{\~n}ados}, {Bell},
  {Bender}, {Bernard}, {Boegner}, {Boffi}, {Botticella}, {Calamida},
  {Casertano}, {Chen}, {Chen}, {Cole}, {Deacon}, {Frenk}, {Fitzsimmons},
  {Gezari}, {Gibbs}, {Goessl}, {Goggia}, {Gourgue}, {Goldman}, {Grant},
  {Grebel}, {Hambly}, {Hasinger}, {Heavens}, {Heckman}, {Henderson}, {Henning},
  {Holman}, {Hopp}, {Ip}, {Isani}, {Jackson}, {Keyes}, {Koekemoer}, {Kotak},
  {Le}, {Liska}, {Long}, {Lucey}, {Liu}, {Martin}, {Masci}, {McLean}, {Mindel},
  {Misra}, {Morganson}, {Murphy}, {Obaika}, {Narayan}, {Nieto-Santisteban},
  {Norberg}, {Peacock}, {Pier}, {Postman}, {Primak}, {Rae}, {Rai}, {Riess},
  {Riffeser}, {Rix}, {R{\"o}ser}, {Russel}, {Rutz}, {Schilbach}, {Schultz},
  {Scolnic}, {Strolger}, {Szalay}, {Seitz}, {Small}, {Smith}, {Soderblom},
  {Taylor}, {Thomson}, {Taylor}, {Thakar}, {Thiel}, {Thilker}, {Unger},
  {Urata}, {Valenti}, {Wagner}, {Walder}, {Walter}, {Watters}, {Werner},
  {Wood-Vasey}, \& {Wyse}}]{chambers2016}
{Chambers}, K.~C., {Magnier}, E.~A., {Metcalfe}, N., {et~al.} 2016, arXiv
  e-prints, arXiv:1612.05560.
\newblock \doarXiv{1612.05560}

\bibitem[{{Damen} {et~al.}(2011){Damen}, {Labb{\'e}}, {van Dokkum}, {Franx},
  {Taylor}, {Brandt}, {Dickinson}, {Gawiser}, {Illingworth}, {Kriek},
  {Marchesini}, {Muzzin}, {Papovich}, \& {Rix}}]{damen2011}
{Damen}, M., {Labb{\'e}}, I., {van Dokkum}, P.~G., {et~al.} 2011, \apj, 727, 1,
  \dodoi{10.1088/0004-637X/727/1/1}

\bibitem[{{Dickinson} {et~al.}(2003){Dickinson}, {Papovich}, {Ferguson}, \&
  {Budav{\'a}ri}}]{Dickinson+03}
{Dickinson}, M., {Papovich}, C., {Ferguson}, H.~C., \& {Budav{\'a}ri}, T. 2003,
  \apj, 587, 25, \dodoi{10.1086/368111}

\bibitem[{{Feldmann} {et~al.}(2017){Feldmann}, {Quataert}, {Hopkins},
  {Faucher-Gigu{\`e}re}, \& {Kere{\v{s}}}}]{feldmann2017}
{Feldmann}, R., {Quataert}, E., {Hopkins}, P.~F., {Faucher-Gigu{\`e}re}, C.-A.,
  \& {Kere{\v{s}}}, D. 2017, \mnras, 470, 1050, \dodoi{10.1093/mnras/stx1120}

\bibitem[{{Forrest} {et~al.}(2018){Forrest}, {Tran}, {Broussard}, {Cohn},
  {Kennicutt}, {Papovich}, {Allen}, {Cowley}, {Glazebrook}, {Kacprzak},
  {Kawinwanichakij}, {Nanayakkara}, {Salmon}, {Spitler}, \&
  {Straatman}}]{forrest2018}
{Forrest}, B., {Tran}, K.-V.~H., {Broussard}, A., {et~al.} 2018, \apj, 863,
  131, \dodoi{10.3847/1538-4357/aad232}

\bibitem[{{Hill} {et~al.}(2017){Hill}, {Muzzin}, {Franx}, {Clauwens},
  {Schreiber}, {Marchesini}, {Stefanon}, {Labbe}, {Brammer}, {Caputi}, {Fynbo},
  {Milvang-Jensen}, {Skelton}, {van Dokkum}, \& {Whitaker}}]{hill2017}
{Hill}, A.~R., {Muzzin}, A., {Franx}, M., {et~al.} 2017, \apj, 837, 147,
  \dodoi{10.3847/1538-4357/aa61fe}

\bibitem[{{Jarvis} {et~al.}(2013){Jarvis}, {Bonfield}, {Bruce}, {Geach},
  {McAlpine}, {McLure}, {Gonz{\'a}lez-Solares}, {Irwin}, {Lewis}, {Yoldas},
  {Andreon}, {Cross}, {Emerson}, {Dalton}, {Dunlop}, {Hodgkin}, {Le},
  {Karouzos}, {Meisenheimer}, {Oliver}, {Rawlings}, {Simpson}, {Smail},
  {Smith}, {Sullivan}, {Sutherland}, {White}, \& {Zwart}}]{jarvis2013}
{Jarvis}, M.~J., {Bonfield}, D.~G., {Bruce}, V.~A., {et~al.} 2013, \mnras, 428,
  1281, \dodoi{10.1093/mnras/sts118}

\bibitem[{{Kim} {et~al.}(2011){Kim}, {Dunlop}, {Lonsdale}, {Farrah}, {Lacy},
  {Sun}, \& {SpUDS Team}}]{kim2011}
{Kim}, M., {Dunlop}, J.~S., {Lonsdale}, C.~J., {et~al.} 2011, in American
  Astronomical Society Meeting Abstracts, Vol. 217, American Astronomical
  Society Meeting Abstracts \#217, 335.51

\bibitem[{{Krefting} {et~al.}(2020){Krefting}, {Sajina}, {Lacy}, {Nyland },
  {Farrah}, {Darvish}, {Duivenvoorden}, {Duncan}, {Gonzalez-Perez}, {del P.
  Lagos}, {Oliver}, {Shirley}, \& {Vaccari}}]{Krefting2020}
{Krefting}, N., {Sajina}, A., {Lacy}, M., {et~al.} 2020, \apj, 889, 185,
  \dodoi{10.3847/1538-4357/ab60a0}

\bibitem[{{Labb{\'e}} {et~al.}(2005){Labb{\'e}}, {Huang}, {Franx}, {Rudnick},
  {Barmby}, {Daddi}, {van Dokkum}, {Fazio}, {F{\"o}rster Schreiber},
  {Moorwood}, {Rix}, {R{\"o}ttgering}, {Trujillo}, \& {van der
  Werf}}]{labbe2005}
{Labb{\'e}}, I., {Huang}, J., {Franx}, M., {et~al.} 2005, \apjl, 624, L81,
  \dodoi{10.1086/430700}

\bibitem[{{Labb{\'e}} {et~al.}(2010){Labb{\'e}}, {Gonz{\'a}lez}, {Bouwens},
  {Illingworth}, {Oesch}, {van Dokkum}, {Carollo}, {Franx}, {Stiavelli},
  {Trenti}, {Magee}, \& {Kriek}}]{labbe2010}
{Labb{\'e}}, I., {Gonz{\'a}lez}, V., {Bouwens}, R.~J., {et~al.} 2010, \apjl,
  708, L26, \dodoi{10.1088/2041-8205/708/1/L26}

\bibitem[{{Labb{\'e}} {et~al.}(2013){Labb{\'e}}, {Oesch}, {Bouwens},
  {Illingworth}, {Magee}, {Gonz{\'a}lez}, {Carollo}, {Franx}, {Trenti}, {van
  Dokkum}, \& {Stiavelli}}]{labbe2013}
{Labb{\'e}}, I., {Oesch}, P.~A., {Bouwens}, R.~J., {et~al.} 2013, \apjl, 777,
  L19, \dodoi{10.1088/2041-8205/777/2/L19}

\bibitem[{{Labb{\'e}} {et~al.}(2015){Labb{\'e}}, {Oesch}, {Illingworth}, {van
  Dokkum}, {Bouwens}, {Franx}, {Carollo}, {Trenti}, {Holden}, {Smit},
  {Gonz{\'a}lez}, {Magee}, {Stiavelli}, \& {Stefanon}}]{labbe2015}
{Labb{\'e}}, I., {Oesch}, P.~A., {Illingworth}, G.~D., {et~al.} 2015, \apjs,
  221, 23, \dodoi{10.1088/0067-0049/221/2/23}

\bibitem[{{Labbe} {et~al.}(2016){Labbe}, {Caputi}, {McLeod}, {Cowley}, {Dayal},
  {Behroozi}, {Ashby}, {Franx}, {Dunlop}, {Le Fevre}, {Fynbo}, {McCracken},
  {Milvang-Jensen}, {Ilbert}, {Tasca}, {de Barros}, {Oesch}, {Bouwens},
  {Muzzin}, {Illingworth}, {Stefanon}, {Schreiber}, {Hutter}, \& {van
  Dokkum}}]{labbe2016}
{Labbe}, I., {Caputi}, K., {McLeod}, D., {et~al.} 2016, {Completing the Legacy
  of Spitzer/IRAC over COSMOS}, Spitzer Proposal ID 13094

\bibitem[{{Lacy} {et~al.}(2021){Lacy}, {Surace}, {Farrah}, {Nyland}, {Afonso},
  {Brandt}, {Clements}, {Lagos}, {Maraston}, {Pforr}, {Sajina}, {Sako},
  {Vaccari}, {Wilson}, {Ballantyne}, {Barkhouse}, {Brunner}, {Cane}, {Clarke},
  {Cooper}, {Cooray}, {Covone}, {D'Andrea}, {Evrard}, {Ferguson}, {Frieman},
  {Gonzalez-Perez}, {Gupta}, {Hatziminaoglou}, {Huang}, {Jagannathan},
  {Jarvis}, {Jones}, {Kimball}, {Lidman}, {Lubin}, {Marchetti}, {Martini},
  {McMahon}, {Mei}, {Messias}, {Murphy}, {Newman}, {Nichol}, {Norris},
  {Oliver}, {Perez-Fournon}, {Peters}, {Pierre}, {Polisensky}, {Richards},
  {Ridgway}, {R{\"o}ttgering}, {Seymour}, {Shirley}, {Somerville}, {Strauss},
  {Suntzeff}, {Thorman}, {van Kampen}, {Verma}, {Wechsler}, \&
  {Wood-Vasey}}]{Lacy2021}
{Lacy}, M., {Surace}, J.~A., {Farrah}, D., {et~al.} 2021, \mnras, 501, 892,
  \dodoi{10.1093/mnras/staa3714}

\bibitem[{{Laporte} {et~al.}(2021){Laporte}, {Meyer}, {Ellis}, {Robertson},
  {Chisholm}, \& {Roberts-Borsani}}]{laporte2021}
{Laporte}, N., {Meyer}, R.~A., {Ellis}, R.~S., {et~al.} 2021, \mnras, 505,
  3336, \dodoi{10.1093/mnras/stab1239}

\bibitem[{{Laureijs} {et~al.}(2011){Laureijs}, {Amiaux}, {Arduini},
  {Augu{\`e}res}, {Brinchmann}, {Cole}, {Cropper}, {Dabin}, {Duvet}, {Ealet},
  {Garilli}, {Gondoin}, {Guzzo}, {Hoar}, {Hoekstra}, {Holmes}, {Kitching},
  {Maciaszek}, {Mellier}, {Pasian}, {Percival}, {Rhodes}, {Saavedra Criado},
  {Sauvage}, {Scaramella}, {Valenziano}, {Warren}, {Bender}, {Castander},
  {Cimatti}, {Le F{\`e}vre}, {Kurki-Suonio}, {Levi}, {Lilje}, {Meylan},
  {Nichol}, {Pedersen}, {Popa}, {Rebolo Lopez}, {Rix}, {Rottgering},
  {Zeilinger}, {Grupp}, {Hudelot}, {Massey}, {Meneghetti}, {Miller}, {Paltani},
  {Paulin-Henriksson}, {Pires}, {Saxton}, {Schrabback}, {Seidel}, {Walsh},
  {Aghanim}, {Amendola}, {Bartlett}, {Baccigalupi}, {Beaulieu}, {Benabed},
  {Cuby}, {Elbaz}, {Fosalba}, {Gavazzi}, {Helmi}, {Hook}, {Irwin}, {Kneib},
  {Kunz}, {Mannucci}, {Moscardini}, {Tao}, {Teyssier}, {Weller}, {Zamorani},
  {Zapatero Osorio}, {Boulade}, {Foumond}, {Di Giorgio}, {Guttridge}, {James},
  {Kemp}, {Martignac}, {Spencer}, {Walton}, {Bl{\"u}mchen}, {Bonoli},
  {Bortoletto}, {Cerna}, {Corcione}, {Fabron}, {Jahnke}, {Ligori}, {Madrid},
  {Martin}, {Morgante}, {Pamplona}, {Prieto}, {Riva}, {Toledo}, {Trifoglio},
  {Zerbi}, {Abdalla}, {Douspis}, {Grenet}, {Borgani}, {Bouwens}, {Courbin},
  {Delouis}, {Dubath}, {Fontana}, {Frailis}, {Grazian}, {Koppenh{\"o}fer},
  {Mansutti}, {Melchior}, {Mignoli}, {Mohr}, {Neissner}, {Noddle}, {Poncet},
  {Scodeggio}, {Serrano}, {Shane}, {Starck}, {Surace}, {Taylor},
  {Verdoes-Kleijn}, {Vuerli}, {Williams}, {Zacchei}, {Altieri}, {Escudero
  Sanz}, {Kohley}, {Oosterbroek}, {Astier}, {Bacon}, {Bardelli}, {Baugh},
  {Bellagamba}, {Benoist}, {Bianchi}, {Biviano}, {Branchini}, {Carbone},
  {Cardone}, {Clements}, {Colombi}, {Conselice}, {Cresci}, {Deacon}, {Dunlop},
  {Fedeli}, {Fontanot}, {Franzetti}, {Giocoli}, {Garcia-Bellido}, {Gow},
  {Heavens}, {Hewett}, {Heymans}, {Holland}, {Huang}, {Ilbert}, {Joachimi},
  {Jennins}, {Kerins}, {Kiessling}, {Kirk}, {Kotak}, {Krause}, {Lahav}, {van
  Leeuwen}, {Lesgourgues}, {Lombardi}, {Magliocchetti}, {Maguire}, {Majerotto},
  {Maoli}, {Marulli}, {Maurogordato}, {McCracken}, {McLure}, {Melchiorri},
  {Merson}, {Moresco}, {Nonino}, {Norberg}, {Peacock}, {Pello}, {Penny},
  {Pettorino}, {Di Porto}, {Pozzetti}, {Quercellini}, {Radovich}, {Rassat},
  {Roche}, {Ronayette}, {Rossetti}, {Sartoris}, {Schneider}, {Semboloni},
  {Serjeant}, {Simpson}, {Skordis}, {Smadja}, {Smartt}, {Spano}, {Spiro},
  {Sullivan}, {Tilquin}, {Trotta}, {Verde}, {Wang}, {Williger}, {Zhao},
  {Zoubian}, \& {Zucca}}]{Laureijs2011}
{Laureijs}, R., {Amiaux}, J., {Arduini}, S., {et~al.} 2011, arXiv e-prints,
  arXiv:1110.3193.
\newblock \doarXiv{1110.3193}

\bibitem[{{Lawrence} {et~al.}(2007){Lawrence}, {Warren}, {Almaini}, {Edge},
  {Hambly}, {Jameson}, {Lucas}, {Casali}, {Adamson}, {Dye}, {Emerson},
  {Foucaud}, {Hewett}, {Hirst}, {Hodgkin}, {Irwin}, {Lodieu}, {McMahon},
  {Simpson}, {Smail}, {Mortlock}, \& {Folger}}]{Lawrence2007}
{Lawrence}, A., {Warren}, S.~J., {Almaini}, O., {et~al.} 2007, \mnras, 379,
  1599, \dodoi{10.1111/j.1365-2966.2007.12040.x}

\bibitem[{{Lindegren} {et~al.}(2018){Lindegren}, {Hern{\'a}ndez}, {Bombrun},
  {Klioner}, {Bastian}, {Ramos-Lerate}, {de Torres}, {Steidelm{\"u}ller},
  {Stephenson}, {Hobbs}, {Lammers}, {Biermann}, {Geyer}, {Hilger}, {Michalik},
  {Stampa}, {McMillan}, {Casta{\~n}eda}, {Clotet}, {Comoretto}, {Davidson},
  {Fabricius}, {Gracia}, {Hambly}, {Hutton}, {Mora}, {Portell}, {van Leeuwen},
  {Abbas}, {Abreu}, {Altmann}, {Andrei}, {Anglada}, {Balaguer-N{\'u}{\~n}ez},
  {Barache}, {Becciani}, {Bertone}, {Bianchi}, {Bouquillon}, {Bourda},
  {Br{\"u}semeister}, {Bucciarelli}, {Busonero}, {Buzzi}, {Cancelliere},
  {Carlucci}, {Charlot}, {Cheek}, {Crosta}, {Crowley}, {de Bruijne}, {de
  Felice}, {Drimmel}, {Esquej}, {Fienga}, {Fraile}, {Gai}, {Garralda},
  {Gonz{\'a}lez-Vidal}, {Guerra}, {Hauser}, {Hofmann}, {Holl}, {Jordan},
  {Lattanzi}, {Lenhardt}, {Liao}, {Licata}, {Lister}, {L{\"o}ffler},
  {Marchant}, {Martin-Fleitas}, {Messineo}, {Mignard}, {Morbidelli}, {Poggio},
  {Riva}, {Rowell}, {Salguero}, {Sarasso}, {Sciacca}, {Siddiqui}, {Smart},
  {Spagna}, {Steele}, {Taris}, {Torra}, {van Elteren}, {van Reeven}, \&
  {Vecchiato}}]{lindegren2018}
{Lindegren}, L., {Hern{\'a}ndez}, J., {Bombrun}, A., {et~al.} 2018, \aap, 616,
  A2, \dodoi{10.1051/0004-6361/201832727}

\bibitem[{{Lonsdale} {et~al.}(2003){Lonsdale}, {Smith}, {Rowan-Robinson},
  {Surace}, {Shupe}, {Xu}, {Oliver}, {Padgett}, {Fang}, {Conrow},
  {Franceschini}, {Gautier}, {Griffin}, {Hacking}, {Masci}, {Morrison},
  {O'Linger}, {Owen}, {P{\'e}rez-Fournon}, {Pierre}, {Puetter}, {Stacey},
  {Castro}, {Polletta}, {Farrah}, {Jarrett}, {Frayer}, {Siana}, {Babbedge},
  {Dye}, {Fox}, {Gonzalez-Solares}, {Salaman}, {Berta}, {Condon}, {Dole}, \&
  {Serjeant}}]{lonsdale2003}
{Lonsdale}, C.~J., {Smith}, H.~E., {Rowan-Robinson}, M., {et~al.} 2003, \pasp,
  115, 897, \dodoi{10.1086/376850}

\bibitem[{{Madau} \& {Dickinson}(2014)}]{madau2014}
{Madau}, P., \& {Dickinson}, M. 2014, \araa, 52, 415,
  \dodoi{10.1146/annurev-astro-081811-125615}

\bibitem[{{Mauduit} {et~al.}(2012){Mauduit}, {Lacy}, {Farrah}, {Surace},
  {Jarvis}, {Oliver}, {Maraston}, {Vaccari}, {Marchetti}, {Zeimann},
  {Gonz{\'a}les-Solares}, {Pforr}, {Petric}, {Henriques}, {Thomas}, {Afonso},
  {Rettura}, {Wilson}, {Falder}, {Geach}, {Huynh}, {Norris}, {Seymour},
  {Richards}, {Stanford}, {Alexand er}, {Becker}, {Best}, {Bizzocchi},
  {Bonfield}, {Castro}, {Cava}, {Chapman}, {Christopher}, {Clements}, {Covone},
  {Dubois}, {Dunlop}, {Dyke}, {Edge}, {Ferguson}, {Foucaud}, {Franceschini},
  {Gal}, {Grant}, {Grossi}, {Hatziminaoglou}, {Hickey}, {Hodge}, {Huang},
  {Ivison}, {Kim}, {LeFevre}, {Lehnert}, {Lonsdale}, {Lubin}, {McLure},
  {Messias}, {Mart{\'\i}nez-Sansigre}, {Mortier}, {Nielsen}, {Ouchi}, {Parish},
  {Perez-Fournon}, {Pierre}, {Rawlings}, {Readhead}, {Ridgway}, {Rigopoulou},
  {Romer}, {Rosebloom}, {Rottgering}, {Rowan-Robinson}, {Sajina}, {Simpson},
  {Smail}, {Squires}, {Stevens}, {Taylor}, {Trichas}, {Urrutia}, {van Kampen},
  {Verma}, \& {Xu}}]{mauduit2012}
{Mauduit}, J.~C., {Lacy}, M., {Farrah}, D., {et~al.} 2012, \pasp, 124, 714,
  \dodoi{10.1086/666945}

\bibitem[{{Merlin} {et~al.}(2018){Merlin}, {Fontana}, {Castellano}, {Santini},
  {Torelli}, {Boutsia}, {Wang}, {Grazian}, {Pentericci}, {Schreiber}, {Ciesla},
  {McLure}, {Derriere}, {Dunlop}, \& {Elbaz}}]{merlin2018}
{Merlin}, E., {Fontana}, A., {Castellano}, M., {et~al.} 2018, \mnras, 473,
  2098, \dodoi{10.1093/mnras/stx2385}

\bibitem[{{Miyazaki} {et~al.}(1998){Miyazaki}, {Sekiguchi}, {Imi}, {Okada},
  {Nakata}, \& {Komiyama}}]{miyazaki1998}
{Miyazaki}, S., {Sekiguchi}, M., {Imi}, K., {et~al.} 1998, in Society of
  Photo-Optical Instrumentation Engineers (SPIE) Conference Series, Vol. 3355,
  Optical Astronomical Instrumentation, ed. S.~{D'Odorico}, 363--374

\bibitem[{{Muzzin} {et~al.}(2009){Muzzin}, {Wilson}, {Yee}, {Hoekstra},
  {Gilbank}, {Surace}, {Lacy}, {Blindert}, {Majumdar}, {Demarco}, {Gardner},
  {Gladders}, \& {Lonsdale}}]{muzzin2009}
{Muzzin}, A., {Wilson}, G., {Yee}, H.~K.~C., {et~al.} 2009, \apj, 698, 1934,
  \dodoi{10.1088/0004-637X/698/2/1934}

\bibitem[{{Nyland} {et~al.}(2017){Nyland}, {Lacy}, {Sajina}, {Pforr}, {Farrah},
  {Wilson}, {Surace}, {H{\"a}u{\ss}ler}, {Vaccari}, \& {Jarvis}}]{Nyland2017}
{Nyland}, K., {Lacy}, M., {Sajina}, A., {et~al.} 2017, \apjs, 230, 9,
  \dodoi{10.3847/1538-4365/aa6fed}

\bibitem[{{Sanders} {et~al.}(2007){Sanders}, {Salvato}, {Aussel}, {Ilbert},
  {Scoville}, {Surace}, {Frayer}, {Sheth}, {Helou}, {Brooke}, {Bhattacharya},
  {Yan}, {Kartaltepe}, {Barnes}, {Blain}, {Calzetti}, {Capak}, {Carilli},
  {Carollo}, {Comastri}, {Daddi}, {Ellis}, {Elvis}, {Fall}, {Franceschini},
  {Giavalisco}, {Hasinger}, {Impey}, {Koekemoer}, {Le F{\`e}vre}, {Lilly},
  {Liu}, {McCracken}, {Mobasher}, {Renzini}, {Rich}, {Schinnerer}, {Shopbell},
  {Taniguchi}, {Thompson}, {Urry}, \& {Williams}}]{sanders2007}
{Sanders}, D.~B., {Salvato}, M., {Aussel}, H., {et~al.} 2007, \apjs, 172, 86,
  \dodoi{10.1086/517885}

\bibitem[{{Satyapal} {et~al.}(2017){Satyapal}, {Secrest}, {Ricci}, {Ellison},
  {Rothberg}, {Blecha}, {Constantin}, {Gliozzi}, {McNulty}, \&
  {Ferguson}}]{Satyapal2017}
{Satyapal}, S., {Secrest}, N.~J., {Ricci}, C., {et~al.} 2017, \apj, 848, 126,
  \dodoi{10.3847/1538-4357/aa88ca}

\bibitem[{{Sawicki} {et~al.}(2019){Sawicki}, {Arnouts}, {Huang}, {Coupon},
  {Golob}, {Gwyn}, {Foucaud}, {Moutard}, {Iwata}, {Liu}, {Chen}, {Desprez},
  {Harikane}, {Ono}, {Strauss}, {Tanaka}, {Thibert}, {Balogh}, {Bundy},
  {Chapman}, {Gunn}, {Hsieh}, {Ilbert}, {Jing}, {LeF{\`e}vre}, {Li}, {Matsuda},
  {Miyazaki}, {Nagao}, {Nishizawa}, {Ouchi}, {Shimasaku}, {Silverman}, {de la
  Torre}, {Tresse}, {Wang}, {Willott}, {Yamada}, {Yang}, \&
  {Yee}}]{sawicki2019}
{Sawicki}, M., {Arnouts}, S., {Huang}, J., {et~al.} 2019, \mnras, 489, 5202,
  \dodoi{10.1093/mnras/stz2522}

\bibitem[{{Scarlata} {et~al.}(2019){Scarlata}, {Capak}, {Finkelstein},
  {Arendt}, {Mehta}, {Dickinson}, {Bruton}, {Hemmati}, {Teplitz}, {Davidzon},
  {Salvato}, {Sanders}, {Scaramella}, {Gardner}, {Hildebrandt}, {Oesch},
  {Brammer}, {Mei}, {McCraken}, {Foley}, {Dole}, {Rhodes}, {Conselice}, {Toft},
  {Ilbert}, {Scolnic}, {Bagley}, {Cuillandre}, {Larson}, {Maraston}, {Cuby},
  {Baronchelli}, \& {Kashlinsky}}]{Scarlata2019}
{Scarlata}, C., {Capak}, P., {Finkelstein}, S., {et~al.} 2019, {The Euclid Deep
  Field South}, Spitzer Proposal

\bibitem[{{Scoville} {et~al.}(2007){Scoville}, {Aussel}, {Brusa}, {Capak},
  {Carollo}, {Elvis}, {Giavalisco}, {Guzzo}, {Hasinger}, {Impey}, {Kneib},
  {LeFevre}, {Lilly}, {Mobasher}, {Renzini}, {Rich}, {Sanders}, {Schinnerer},
  {Schminovich}, {Shopbell}, {Taniguchi}, \& {Tyson}}]{scoville2007}
{Scoville}, N., {Aussel}, H., {Brusa}, M., {et~al.} 2007, \apjs, 172, 1,
  \dodoi{10.1086/516585}

\bibitem[{{Sherman} {et~al.}(2021){Sherman}, {Jogee}, {Florez}, {Finkelstein},
  {Ciardullo}, {Wold}, {Stevans}, {Kawinwanichakij}, {Papovich}, \&
  {Gronwall}}]{sherman2021}
{Sherman}, S., {Jogee}, S., {Florez}, J., {et~al.} 2021, \mnras, 505, 947,
  \dodoi{10.1093/mnras/stab1350}

\bibitem[{{Stefanon} {et~al.}(2022){Stefanon}, {Bouwens}, {Labb{\'e}},
  {Illingworth}, {Gonzalez}, \& {Oesch}}]{stefanon2022b}
{Stefanon}, M., {Bouwens}, R.~J., {Labb{\'e}}, I., {et~al.} 2022, arXiv
  e-prints, arXiv:2206.13525.
\newblock \doarXiv{2206.13525}

\bibitem[{{Stefanon} {et~al.}(2018){Stefanon}, {Labbe}, {Caputi}, {Bouwens},
  {Oesch}, {Ashby}, {Dunlop}, {Franx}, {Fynbo}, {Illingworth}, {Le Fevre},
  {Marchesini}, {McCracken}, {Milvang Jensen}, {Muzzin}, \& {van
  Dokkum}}]{stefanon2018}
{Stefanon}, M., {Labbe}, I., {Caputi}, K., {et~al.} 2018, {COMPLETE2:
  Completing the Legacy of Spitzer/IRAC over COSMOS}, Spitzer Proposal ID
  \#14045

\bibitem[{{Stefanon} {et~al.}(2021){Stefanon}, {Labb{\'e}}, {Oesch}, {De
  Barros}, {Gonzalez}, {Bouwens}, {Franx}, {Illingworth}, {Holden}, {Magee},
  {Smit}, \& {van Dokkum}}]{stefanon2021}
{Stefanon}, M., {Labb{\'e}}, I., {Oesch}, P.~A., {et~al.} 2021, \apjs, 257, 68,
  \dodoi{10.3847/1538-4365/ac2498}

\bibitem[{{Steinhardt} {et~al.}(2014){Steinhardt}, {Speagle}, {Capak},
  {Silverman}, {Carollo}, {Dunlop}, {Hashimoto}, {Hsieh}, {Ilbert}, {Le Fevre},
  {Le Floc'h}, {Lee}, {Lin}, {Lin}, {Masters}, {McCracken}, {Nagao}, {Petric},
  {Salvato}, {Sanders}, {Scoville}, {Sheth}, {Strauss}, \&
  {Taniguchi}}]{steinhardt2014}
{Steinhardt}, C.~L., {Speagle}, J.~S., {Capak}, P., {et~al.} 2014, \apj, 791,
  L25, \dodoi{10.1088/2041-8205/791/2/L25}

\bibitem[{{Strait} {et~al.}(2020){Strait}, {Brada{\v{c}}}, {Coe}, {Bradley},
  {Salmon}, {Lemaux}, {Huang}, {Zitrin}, {Sharon}, {Acebron}, {Andrade-Santos},
  {Avila}, {Frye}, {Hoag}, {Mahler}, {Nonino}, {Ogaz}, {Oguri}, {Ouchi},
  {Paterno-Mahler}, \& {Pelliccia}}]{strait2020}
{Strait}, V., {Brada{\v{c}}}, M., {Coe}, D., {et~al.} 2020, \apj, 888, 124,
  \dodoi{10.3847/1538-4357/ab5daf}

\bibitem[{{Tanaka} {et~al.}(2017){Tanaka}, {Hasinger}, {Silverman},
  {Bickerton}, {Furusawa}, {Harikane}, {Hu}, {Ikeda}, {Li}, {McCracken},
  {Price}, {Strauss}, {Koike}, {Komiyama}, {Mineo}, {Miyazaki}, {Nishizawa},
  {Takata}, {Utsumi}, {Yamada}, \& {Yasuda}}]{tanaka2017}
{Tanaka}, M., {Hasinger}, G., {Silverman}, J.~D., {et~al.} 2017, ArXiv
  e-prints.
\newblock \doarXiv{1706.00566}

\bibitem[{{Timlin} {et~al.}(2016){Timlin}, {Ross}, {Richards}, {Lacy}, {Ryan},
  {Stone}, {Bauer}, {Brandt}, {Fan}, {Glikman}, {Haggard}, {Jiang}, {LaMassa},
  {Lin}, {Makler}, {McGehee}, {Myers}, {Schneider}, {Urry}, {Wollack}, \&
  {Zakamska}}]{timlin2016}
{Timlin}, J.~D., {Ross}, N.~P., {Richards}, G.~T., {et~al.} 2016, \apjs, 225,
  1, \dodoi{10.3847/0067-0049/225/1/1}

\bibitem[{{Torra} {et~al.}(2021){Torra}, {Casta{\~n}eda}, {Fabricius},
  {Lindegren}, {Clotet}, {Gonz{\'a}lez-Vidal}, {Bartolom{\'e}}, {Bastian},
  {Bernet}, {Biermann}, {Garralda}, {G{\'u}rpide}, {Lammers}, {Portell}, \&
  {Torra}}]{torra2021}
{Torra}, F., {Casta{\~n}eda}, J., {Fabricius}, C., {et~al.} 2021, \aap, 649,
  A10, \dodoi{10.1051/0004-6361/202039637}

\bibitem[{{Whitaker} {et~al.}(2013){Whitaker}, {van Dokkum}, {Brammer},
  {Momcheva}, {Skelton}, {Franx}, {Kriek}, {Labb{\'e}}, {Fumagalli},
  {Lundgren}, {Nelson}, {Patel}, \& {Rix}}]{whitaker2013}
{Whitaker}, K.~E., {van Dokkum}, P.~G., {Brammer}, G., {et~al.} 2013, \apjl,
  770, L39, \dodoi{10.1088/2041-8205/770/2/L39}

\end{thebibliography}

\end{document}